\documentclass[11pt, letterpaper]{article}
\usepackage{fullpage}
\usepackage{amsmath}
\usepackage{amssymb}
\usepackage{hyperref}
\usepackage{xcolor}
\usepackage{parskip}
\usepackage{graphicx}
\usepackage{subcaption}
\newcommand{\nn}{\nonumber}
\newcommand{\ba}{\begin{align}}
\newcommand{\ea}{\end{align}}
\newcommand{\be}{\begin{equation}}
\newcommand{\ee}{\end{equation}}
\newcommand{\bea}{\begin{eqnarray}}
\newcommand{\eea}{\end{eqnarray}}
\newcommand{\bp}{\begin{pmatrix}}
\newcommand{\ep}{\end{pmatrix}}
\newcommand{\bi}{\begin{itemize}}
\newcommand{\ei}{\end{itemize}}
\newcommand{\overbar}[1]{\mkern 1.5mu\overline{\mkern-1.5mu#1\mkern-1.5mu}\mkern 1.5mu}
\newcommand{\m}{\mathcal}

\newcommand{\D}{\mathcal{D}}
\newcommand{\oD}{\overbar{\mathcal{D}}}
\newcommand{\p}{\partial}
\newcommand{\La}{\mathcal{L}}

\newcommand{\hpi}{\hat{\pi}}

\newcommand{\M}{M_{P}^2}
\newcommand{\K}{\tfrac{K}{\M}}

\newcommand{\Km}{\tfrac{K}{\M}}
\newcommand{\x}{\frac{K}{M_P^2}}

\begin{document}

\bigskip
\thispagestyle{empty}
\bigskip
\begin{center}
{\LARGE \bf $N=1$ Supergravitational Heterotic Galileons}
\end{center}

\bigskip
\bigskip
\begin{center}
{\large {\bf Rehan Deen} and {\bf Burt Ovrut}}
\end{center}

\begin{center}
    {\it Department of Physics and Astronomy\\
     University of Pennsylvania \\
     Philadelphia, PA 19104--6396}\\
\end{center}

\bigskip
\bigskip
\bigskip
\begin{center}
\textbf{Abstract}\end{center}
{\noindent
Heterotic $M$-theory consists of a five-dimensional manifold of the form $S^1/\mathbf{Z}_2 \times M_{4}$. It has been shown that one of the two orbifold planes, the ``observable'' sector, can have a low energy particle spectrum which is precisely the $N=1$ supersymmetric standard model with three right-handed neutrino chiral supermultiplets. The other orbifold plane constitutes a ``hidden'' sector which, since its communication with the observable sector is suppressed, will be ignored in this paper. However, the finite fifth-dimension allows for the existence of three-brane solitons which, in order to render the vacuum anomaly free, must appear. That is, heterotic $M$-theory provides a natural framework for brane-world cosmological scenarios coupled to realistic particle physics.
The complete worldvolume action of such three-branes is unknown. Here, treating these solitons as probe branes, we 
construct their scalar worldvolume Lagrangian as a derivative expansion of the heterotic DBI action. In analogy with similar calculations in the $M_{5}$ and $AdS_{5}$ context, this leads to the construction of ``heterotic Galileons''. However, realistic vacua of heterotic $M$-theory are necessarily $N=1$ supersymmetric in four dimensions. Hence, we proceed to supersymmetrize the three-brane worldvolume action, first in flat superspace and then extend the results to $N=1$ supergravity. Such a worldvolume action may lead to interesting cosmology, such as ``bouncing" universe models, by allowing for the violation of the Null Energy Condition (NEC).}

\vskip 7cm
\hrulefill

Email:~~rdeen@sas.upenn.edu,~ ovrut@elcapitan.hep.upenn.edu

\newpage 


\section{Introduction}


The relatively recent discovery of the Higgs boson \cite{Aad:2012tfa, Chatrchyan:2012xdj} showed that low energy particle physics is apparently described, to a high degree of accuracy, by the so-called ``standard model''. However, it also had a potentially important impact on theories of early universe cosmology. Specifically, it demonstrated the existence in nature of spin-0 fields, required by particle physics phenomenology, that could potentially act as the dynamical cosmological scalar. In \cite{Bezrukov:2007ep}, an attempt was made to use the Higgs boson as a natural ``inflaton'' within the inflationary scenario. Although compelling, this theory was plagued by serious problems--such as the requirement that the Higgs field be quadratically coupled, with an unnaturally large coupling parameter, to the curvature scalar in the Lagrangian density. Attempts to circumvent these issues by, for example, extending the standard model and, hence, the Higgs boson to $N=1$ supersymmetry \cite{Einhorn:2009bh, Ferrara:2010yw} also manifested significant problems. A somewhat different approach, using right-handed sneutrino scalars instead of the Higgs boson as an inflaton within the context of the $N=1$ supersymmetric ``$B-L$ MSSM'' theory \cite{Deen:2016zfr}, was, in principle, more successful. However, it is plagued by all of the initial value and multiverse issues involved in any inflationary scenario. Be that as it may, the concept that any realistic theory of cosmology should 1) contain the standard model of particle physics, 2) naturally introduce the scalar or scalars associated with early universe dynamics and 3) imply the exact scalar self-couplings, as well as their explicit coupling to dynamical four-dimensional gravitation, was made very compelling. Since the various problems plaguing inflationary scenarios seem difficult to overcome, in this paper we will instead consider so-called ``bouncing" theories of cosmology, see \cite{Khoury:2001wf}-\cite{Buchbinder:2007at} and \cite{Creminelli:2007aq}-\cite{Ijjas:2016vtq}. Natural versions of bouncing cosmologies should also satisfy criteria 1), 2) and 3). However, a fourth criterion must be added; namely that 4) the theory naturally allow for the violation of the ``null energy condition" (NEC)--as is required for spacetime to transition from a contracting to an expanding phase. In this paper, we will discuss theories that satisfy all four conditions.

It is well-known that this fourth condition can naturally manifest itself in worldvolume theories of 3+1 dimensional bosonic branes. For example, it was shown in \cite{ Creminelli:2010ba, Hinterbichler:2012yn} that the worldvolume theory of a three-brane embedded in an $AdS_{5}$ bulk space can, for the appropriate choice of coefficients, violate the NEC. It follows that co-dimension one bosonic branes embedded in various five-dimensional bulk spaces are potentially of interest in theories of cosmology. The generic form for the worldvolume action of such branes, subject to the restriction that the associated equations of motion have at most two derivatives, has been presented in \cite{Goon:2011qf, Goon:2011uw} for the maximally symmetric bulk spaces $AdS_{5}$, $dS_{5}$ and $M_{5}$. In these cases, the three-brane Lagrangians are potentially interesting in their own right. For example, the 3+1 dimensional bosonic brane embedded in $AdS_{5}$, when expanded into terms each containing the same number of derivatives, exactly reproduces the so-called ``conformal Galileons'' originally presented in \cite{Dvali:2000hr,Nicolis:2008in,deRham:2010eu}. However, none of these theories contain the standard model of particle physics and are not associated with it in any natural way. That is, these theories violate conditions 1) and 2) specified above and, hence, do not constitute fundamental theories of particle cosmology. However, there is a specific bosonic three-brane worldvolume theory that does not suffer from these drawbacks. This theory consists of a co-dimension one three-brane embedded in the  five-dimensional bulk space $S_{1}/\mathbf{Z}_{2} \times M_{4}$. As discussed in detail below, this arises as part of the low energy vacuum--called ``heterotic $M$-theory'' \cite{Lukas:1997fg, Lukas:1998yy, Lukas:1998tt}--of a compactification of 11-dimensional $M$-theory. In order to set the context for this fundamental cosmological three-brane, as well as for its $N=1$ supersymmetric and supergravitational extensions, a careful discussion of heterotic $M$-theory is presented in Section 2.

The bosonic worldvolume theory of this heterotic brane is constructed in detail in Section 3, using the formalism first presented for the $AdS_{5}$, $dS_{5}$ and $M_{5}$ bulk spaces in \cite{Goon:2011qf, Goon:2011uw}.  The single real scalar field, which we will denote by $\pi$, is a function of the four spacetime coordinates on the brane surface and specifies the exact embedding of the three-brane in the bulk space.  We begin by writing out the complete DBI action for the bosonic three-brane worldvolume using the metric for the five-dimensional heterotic $M$-theory bulk space derived in \cite{Ovrut:2012wn} and specified in Section 2. The dimension one parameter, $\alpha$, that arises in this metric is also defined in Section 2. In particular, the four-dimensional spacetime derivatives $\partial$ enter the action via powers of the dimensionless ratio $(\partial/\alpha)^{2}$. As in the case of the three-brane embedded in $AdS_{5}$ space presented in \cite{Goon:2011qf, Goon:2011uw, Deen:2017jqv}, we perform a ``derivative expansion'' of the heterotic worldvolume action in terms of this dimensionless ratio. Unlike the $AdS_{5}$ case, whose derivative expansion naturally terminates in the five conformal Galileons, the derivative expansion of the bosonic heterotic worldvolume is infinite. However, under the assumption that $\partial \ll \alpha$, required so that the effective heterotic theory remains valid, the higher derivative terms become less and less important. In this paper, for simplicity, we will terminate the derivative expansion at order $(\partial/\alpha)^{6}$. Combining all terms with the same order of derivatives together, we find that this truncated action contains four terms, ${\cal{L}}_{i}$, for $i=1,2,3,4$. The exact form of these Lagrangian densities will be derived in Section 3 and constitute a {\it new class} of Galileons, which we will refer to as the ``heterotic Galileons''. We note that ${\cal{L}}_{1}$, contains no derivatives and is simply a function of the scalar field. This implies that this worldvolume theory has a natural potential energy associated with it. On the other hand, ${\cal{L}}_{2}$ is of order $(\partial/\alpha)^{2}$ and constitutes the scalar field kinetic energy, whereas ${\cal{L}}_{3}$ and ${\cal{L}}_{4}$ are specific higher derivative scalar interactions.

The heterotic three-brane bosonic worldvolume action is of interest in its own right. However, the fact that the three-brane arises within the context of heterotic $M$-theory, requires that the worldvolume action be extended to be $N=1$ supersymmetric--as discussed in Section 2. It follows that the fields describing this action must be generalized from the single scalar field, $\hpi=\alpha\pi$, to the complex scalar $A=\frac{1}{\sqrt{2}}(\hpi+i\chi)$, Weyl fermion $\psi$ and complex auxiliary field $F$ of an $N=1$ chiral supermultiplet. We begin by presenting the flat superspace generalization of the four ``heterotic Galileon'' terms ${\cal{L}}_{i}$, for $i=1,2,3,4$ in the derivative expansion of the DBI bosonic action. These superfield expressions will then be expanded into the supersymmetry component fields. Since it is not necessary for this paper, we will ignore the Weyl fermion, but will expand the theory to all orders in both the scalar field $A$ and the auxiliary field $F$. Although not directly applicable to cosmology, which requires its extension to supergravity, this flat superspace theory will allow us to discuss, and solve, two important and related questions that arise in this context. The first involves the introduction of a ``superpotential'' into the Lagrangian. This constitutes the supersymmetric realization of the scalar  potential energy found in ${\cal{L}}_{1}$. The second involves the question of how to ``eliminate'' the auxiliary field $F$ in higher-derivative supersymmetric theories. Both of these issues were discussed in \cite{Deen:2017jqv} and will be resolved in the heterotic context in this paper. Inserting the solution for the $F$ field back into the theory will lead to the complete flat supersymmetric expression for the heterotic worldvolume action.  These flat supersymmetric Lagrangians will be denoted by ${\bar{\cal{L}}}_{i}$ and the results, both in superfields and component fields, will be presented in Section 4.

In order to be relevant to cosmology, it is necessary to generalize these flat superspace results to $N=1$ supergravitation. That is, one must couple the chiral superfield discussed in Section 3 to the $N=1$ supergravity multiplet consisting of the four-dimensional metric $g_{\mu\nu}$, its fermionic gravitino superpartner $\psi^{~\alpha}_{\mu}$, the complex scalar auxiliary field $M$ and the real vector auxiliary field $b_{\mu}$. This is carried out in Section 5 for the first three flat supersymmetric Lagrangians; that is,  ${\bar{\cal{L}}}_{i}$, for $i=1,2,3$. These results will be presented, first, in terms of the associated superfields and second, after the elimination of the supergravity auxiliary fields $M$ and $b_{\mu}$,
in terms of the component fields--again with both fermions $\psi$ and $\psi^{~\alpha}_{\mu}$ being set to zero. We will also present the supergravity extension of ${\bar{\cal{L}}}_{4}$ in terms of superfields. However, the elimination of the supergravity auxiliary fields $M$ and $b_{\mu}$ is much more complicated in the case of ${\bar{\cal{L}}}_{4}$. Hence, in this paper, we will present only that part of the component field expansion of this term not involving these two auxiliary fields. A complete description of the component field Lagrangian associated with ${\bar{\cal{L}}}_{4}$ will be presented elsewhere. However, this is not required in this paper for the following reason. Since we are primarily interested in the cosmological aspects of this heterotic brane worldvolume theory, and since heterotic $M$-theory is only valid for momenta much smaller than the Planck mass $M_{P}$, the Calabi-Yau mass $M_{CY}\simeq 10^{16}$ GeV and the five-dimensional curvature $\alpha \simeq 10^{14}$ GeV, the supergravity extensions of each of ${\bar{\cal{L}}}_{i}$, for $i=1,2,3,4$ need only be evaluated at momenta much smaller than these mass scales. Hence, in ${\bar{\cal{L}}}_{i}$, for $i=1,2,3$ one can ignore all terms suppressed by these scales. The result will reduce to the flat superspace Lagrangians presented in Section 4, where the flat four-dimensional metric $\eta_{\mu\nu}$ is now replaced by a generic metric $g_{\mu\nu}$ and the flat derivative $\partial$ is replaced by the associated covariant derivative $\nabla$. The situation for ${\bar{\cal{L}}}_{4}$ however, is, importantly, somewhat different. It remains true that one takes the flat superspace Lagrangian and replaces $\eta_{\mu\nu} \longrightarrow g_{\mu\nu}$ and $\partial \longrightarrow \nabla$. However, it turns out that there are now two additional terms, {\it not arising from the elimination of $M$ and $b_{\mu}$}, that are purely supergravitational--that is, do not occur in the flat superspace ${\bar{\cal{L}}}_{4}$--and which are {\it not suppressed} by $M_{P}$, $M_{CY}$ or $\alpha$. These two terms involve the curvature tensor associated with $g_{\mu\nu}$ and are explicitly calculated in Section 5. Combining these covariant component field results for ${\bar{\cal{L}}}_{i}$, for $i=1,2,3,4$ leads to the complete $N=1$ supersymmetric heterotic three-brane worldvolume Lagrangian at cosmological energy/momentum scales. This final result will be presented in Section 6. For completeness, in Appendix A we present the entire $N=1$ supersymmetric Lagrangian up to ${\bar{\cal{L}}}_{4}$ in component fields, and in Appendix B we give the $N=1$ supergravitational extension, where we have taken the ``cosmological" limit described above.

It follows that the $N=1$ supergravitational worldvolume theory of a heterotic three-brane explicitly satisfies conditions 2) and 3) discussed above, as well as potentially satisfying condition 4). However, what distinguishes this theory from all of the ``brane worldvolume'' scenarios that preceded it, is that the complete heterotic $M$-theory vacuum in which it naturally arises, explicitly contains the standard model of particle physics--although in its $N=1$ supersymmetric form. That is, the vacua of heterotic $M$-theory naturally satisfy all four conditions for a fundamental theory of particle cosmology.


\section{Heterotic $M$-Theory}


\indent As discussed in the Introduction, in this paper we want to work within the framework of a fundamental theory that a) explicitly contains the exact particle spectrum of the standard model, b) has a natural candidate for the scalar field(s) associated with the cosmological dynamics in the early universe, c) explicitly predicts the interactions of this scalar field(s) with itself and four-dimensional spacetime gravitation and d) allows, in principle, for the violation of the NEC. As we now explain, a compelling choice for this fundamental framework is heterotic $M$-theory \cite{Lukas:1997fg, Lukas:1998yy, Lukas:1998tt}. 

Heterotic M-theory is defined to be the compactification of 11-dimensional Horava-Witten theory \cite{Horava:1995qa, Horava:1996ma} to 5-dimensions. 
This is accomplished in two steps. First, one compactifies Horava-Witten theory on a Calabi-Yau threefold--with or without non-trivial homotopy. This reduces the theory to two 4-dimensional spacetime surfaces, each located at one of the fixed-points of $S_{1}/{\bf{Z}}_{2}$, separated by a finite 5-th dimension. Second, a gauge connection with structure group contained in $E_{8}$, and satisfying the traceless hermitian Yang-Mills equations \cite{Lukas:1998hk, Donagi:1999gc}, is specified on the Calabi-Yau threefold associated with each of these two surfaces. The low energy gauge group and particle spectrum on each 4-dimensional orbifold surface is determined by the choice of this gauge connection \cite{Donagi:2004ia}, as well as by any locally flat ``Wilson lines''. Finally, 
there can be a finite number of codimension-1 ``three branes'' located at various points within the 5-th dimension. These arise from topological five-branes in $M$-theory, each with two spatial dimensions wrapped on a holomorphic curve in the Calabi-Yau threefold, that must satisfy a specific homological constraint \cite{Lukas:1998hk, Donagi:1999jp}.
It is important to note that since 1) the compactification manifold is a Calabi-Yau threefold, 2) the gauge connections each satisfy the traceless hermitian Yang-Mills equations and 3) that every five-brane is wrapped on a holomorphic curve,  the low energy theory on each 4-dimensional orbifold surface, as well as the worldvolume action on each three-brane, must be 
{\it $N=1$ supersymmetric}.

There is clearly a very large number of heterotic $M$-theories that can be constructed, depending on the choice of the Calabi-Yau threefold as well as the specific gauge connections--that is, slope stable holomorphic vector bundles with vanishing slope--chosen on each orbifold surface. However, it was shown in a series of papers \cite{Braun:2004xv, Braun:2005ux, Braun:2005bw, Braun:2005zv, Braun:2005nv} that it is possible to pick both the compactification geometry and as well as the choice of vector bundles so that the low energy physics is phenomenologically realistic. This set of realistic vacua is called the {\it ``heterotic standard model''}. To be specific, the Calabi-Yau threefold is chosen to be a quotient threefold of the form
\begin{equation}
X=\frac{\tilde{X}}{{\bf Z}_{3}\times {\bf Z}_{3}} \ ,
\label{1}
\end{equation}
where 
\begin{equation}
\tilde{X}=dP_{9} \times_{P_{1}} dP_{9}
\label{2}
\end{equation}
is a ``Schoen'' threefold with isometry group ${\bf Z}_{3} \times {\bf Z}_{3}$. In \cite{Braun:2004xv} it was shown that $X$ has three K\"ahler and three complex structure moduli, that is, $h^{1,1}=h^{1,2}=3$, a specific set of intersection numbers $d_{ijk}$ and homotopy group $\pi^{1}={\bf Z}_{3} \times {\bf Z}_{3}$. In the following, let us refer to one of the 4-dimensional orbifold surfaces as the ``observable sector'' and to the other surface as the ``hidden sector''. Then, it was proven in \cite{Braun:2005ux, Braun:2005bw, Braun:2005zv, Braun:2005nv} that one can choose a slope-stable, holomorphic vector bundle with vanishing slope on the observable sector of the form
\begin{equation}
V=\frac{\tilde{V}}{{\bf Z}_{3}\times {\bf Z}_{3}} \ ,
\label{3}
\end{equation}
where $\tilde{V}$ has structure group $SU(4) \subset E_{8}$ and is constructed by ``extension'' as
\begin{equation}
V_{1} \longrightarrow \tilde{V} \longrightarrow V_{2} \ .
\label{4}
\end{equation}
Each of $V_{1}$ and $V_{2}$ is a specific tensor product of a line bundle with a rank two bundle pulled back from a $dP_{9}$ factor of $\tilde{X}$. It was explicitly shown in \cite{Braun:2005zv, Braun:2005nv} that this bundle is ${\bf Z}_{3} \times {\bf Z}_{3}$ ``equivariant'', as it must be. In addition, non-trivial ${\bf Z}_{3} \times {\bf Z}_{3}$ Wilson lines with specific actions on the representations $R$ of $SU(4)$ are introduced. The particle spectrum on the quotient threefold $X$ is obtained by tensoring the cohomology $H^{1}(\tilde{X}, U_{R}(\tilde{V}))$--where $U_{R}(\tilde{V})$ is the tensor product of the bundle associated the $SU(4)$ representation R--with the representation space of $R$ and then taking the ${{\bf Z}_{3} \times {\bf Z}_{3}}$ invariant part.  It was shown in \cite{Braun:2005nv} that the spectrum given by 
\begin{equation}
( H^{1}(\tilde{X}, U_{R}(\tilde{V})\otimes R\big)^{{\bf Z}_{3} \times {\bf Z}_{3}} 
\label{5}
\end{equation}
is {\it exactly} that of the MSSM with three right-handed neutrino chiral super-multiplets--one for each of the three families. Since the ${{\bf Z}_{3} \times {\bf Z}_{3}}$ finite group is Abelian, it follows that the gauge group of the MSSM is 
\begin{equation}
G=SU(3)_{C} \times SU(2)_{L} \times U(1)_{Y} \times U(1)_{B-L} \ .
\label{6}
\end{equation}
That is, it is the standard model gauge group augmented by an extra gauged $U(1)_{B-L}$ factor. We conclude that the heterotic $M$-theory vacuum with this observable sector explicitly satisfies requirement a) above; that is, it {\it explicitly contains the exact particle spectrum of the standard model}--although extended to $N=1$ supersymmetry.

As discussed above, the complete heterotic $M$-theory vacuum requires that one also specify a slope stable, holomorphic vector bundle with vanishing slope on the Calabi-Yau threefold associated with the hidden sector. This choice is far from unique, only being restricted by the requirement that the homological constraint 
\begin{equation}
c_{2}(V^{(observable)})+c_{2}(V^{(hidden)})+W-c_{2}(TX)=0 
\label{7}
\end{equation}
be satisfied. Here, $c_{2}$ specifies the second Chern class and $TX$ is the tangent bundle of the quotient Calabi-Yau threefold $X$. $W$ specifies the homology class associated with the three-branes in the finite 5-th dimensional interval--henceforth, referred to as the ``bulk space''. An explicit example of a hidden sector bundle $V^{(hidden)}$ which, for $X$ and $V$ given in \eqref{1} and \eqref{3} respectively, satisfies condition \eqref{7} for an ``effective'' homology class $W$ is given in \cite{Braun:2013wr}. However, we expect there to be many such hidden sector bundles. Since their spectrum is connected to our observable world only by gravitational suppressed interactions, we, henceforth, ignore the hidden sector. What is {\it vitally important} to this paper, however, is the existence of an effective homology class $W$, which contains holomorphic curves on which two spatial dimensions of a bulk space  five-brane can be wrapped.
We will, henceforth, assume that there is only a {\it single} five-brane wrapped on a holomorphic curve in $W$. That is, our heterotic $M$-theory vacuum contains a single, isolated three-brane in the bulk space. Since the curve is holomorphic, the worldvolume theory of this three-brane must be $N=1$ supersymmetric. The possible intrinsic fields on the three-brane worldvolume were discussed in detail in \cite{Lukas:1998hk}. In general, for a specific gauge choice, the worldvolume contains two real scalar fields--$\pi$, which specifies its embedding in the bulk space and $\chi$, which is the dual to an antisymmetric tensor on the brane surface. These combine together to form a complex ``universal'' scalar, which is the lowest component of a chiral superfield. Additionally, if the genus of the holomorphic curve is $g$, there can also exists $g$ vector superfields on the three-brane worldvolume. Henceforth, for simplicty, we will assume that the holomorphic curve has genus zero and, therefore, these vector supermultiplets do not arise. We conclude that our heterotic $M$-theory vacuum with such a holomorphic curve and the associated bulk space three-brane explicitly satisfies requirement b) above; that is, it {\it has a natural candidate for the scalar field(s) associated with the cosmological dynamics in the early universe}--although extended to $N=1$ supersymmetry.

The main content of this paper will be to explicitly construct the worldvolume theory of this bulk space three-brane; first in flat $N=1$ superspace and then to extend these results to curved $N=1$ supergravity--albeit in the limit where all momenta $\partial \ll \alpha < M_{CY} < M_P$. The result of doing this is two-fold. First, the supergravity extension explicitly satisfies requirement c) above;
that is, it {\it explicitly predicts the interactions of this scalar field(s) with itself and with four-dimensional spacetime gravitation}. The second important consequence of this construction is due to the fact that the worldvolume Lagrangian of this heterotic $M$-theory bulk three-brane contains several terms--each with its own arbitrary parameter. These constants can be constrained to allow for all required properties of the low-energy effective theory--for example, that it be ghost-free. Once this has been done, we expect there to be sufficient freedom left in these coefficients to possibly allow for the violation of the NEC in certain cosmological solutions. This has been shown to be the case in bosonic  three-brane worldvolume theories--such as those involving ``conformal Galileons''--and we expect it to be the case for heterotic $M$-theory bulk space three-branes. Therefore, we expect our heterotic $M$-theory to satisfy condition 4) above; that is, that it {\it allows, in principle, for the violation of the NEC}.

Before continuing to the construction of supersymmetric heterotic three-brane actions, there remains one more important issue that must be discussed; namely, the form of the five-dimensional bulk space metric in heterotic $M$-theory. This was worked out in a number of different contexts in \cite{Lukas:1998tt}. Choosing a flat foliation of the bulk space, the general form of the five-dimensional metric is given by
\begin{equation}
ds^{2}=a(y)^{2} \eta_{\mu\nu} dx^{\mu}dx^{\nu} + b(y)^{2}dy^{2} \ ,
\label{8}
\end{equation}
where $y \in [0,\pi \rho]$ is the coordinate of the finite 5-th dimension and the functions $a(y)$ and $b(y)$ are determined by solving the equations of motion derived from the five-dimensional heterotic $M$-theory Lagrangian. This is straightforward for Calabi-Yau threefold compactifications with $h^{1,1}=1$ \cite{Lukas:1998tt, Brandle:2001ts}. However, for compactifications where $h^{1,1}>1$, this is considerably more difficult. Be that as it may, the solutions for the heterotic standard model, where $h^{1,1}=3$,  were presented in a ``linearized'' approximation in \cite{Lukas:1998tt} . In this case,
{\it assuming there is no three-brane in the bulk space}, one finds
\begin{equation}
a^{2}(y)=a_{0}^{2}h(y)~, ~b^{2}(y)=b_{0}^{2}h(y)^{4} \ ,
\label{9}
\end{equation}
where
\begin{equation}
h(y)=-\frac{2}{3}(\alpha y + c_{0}) 
\label{10}
\end{equation}
and $a_{0}$, $b_{0}$ and $c_{0}$ are dimensionless constants. The dimension one parameter $\alpha$ is defined by
\begin{equation}
\alpha =\frac{\pi}{\sqrt{2}}\bigg( \frac{\kappa}{4\pi}\bigg)^{2/3}\frac{1}{v^{2/3}}\beta \ ,
\label{11}
\end{equation}
with  $\kappa$ the 11-dimensional Planck constant and $v$ is the Calabi-Yau ``reference'' volume, with mass dimensions $-9/2$ and $-6$ respectively,
and 
\begin{equation}
\beta=\frac{1}{v^{1/3}}\int_{X}{\bigg(c_{2}(V^{(observable)})-\frac{1}{2}c_{2}(TX)\bigg)\wedge \omega} \ ,
\label{12}
\end{equation}
where $\omega$ the K\"ahler form on $X$. 
For the case of a single three-brane located at the point $Y \in [0,\pi \rho]$, it was shown in \cite{Brandle:2001ts, Antunes:2002hn} that this solution for $h(y)$ generalizes to 
\begin{equation}
h(y)=-\frac{2}{3}\big( (\alpha + \alpha^{(3)})y -\alpha^{(3)}Y+ c_{0} \big) \ ,
\label{13}
\end{equation}
where the three-brane charge $\alpha^{(3)}$ is 
\begin{equation}
\alpha^{(3)} =\bigg(\frac{\pi}{\sqrt{2}}\bigg( \frac{\kappa}{4\pi}\bigg)^{2/3}\frac{1}{v^{2/3}}\bigg) \frac{1}{v^{1/3}}\int_{X}{W \wedge \omega} \ .
\label{14}
\end{equation}
Here, $W$ is the two-form associated with the wrapped three-brane and satisfies homology condition \eqref{7}. Clearly, the dimension one parameter $\alpha^{(3)}$ depends explicitly on the choice of the hidden sector gauge bundle. For different hidden sector bundles,  $\alpha^{(3)}$ can be either smaller or larger than the observable sector parameter $\alpha$. Since, in this paper, we are ignoring any discussion of the hidden sector, we will simply use the {\it ``probe brane''} approximation; that is, we assume the three-brane does not back-react on the geometry and, hence, does not effect the 5-dimensional metric presented in \eqref{9} and \eqref{10}. We expect this to be a good approximation for certain choices of the hidden sector bundle. In any case, we will, for simplicity, use the ``probe brane'' approximation in the remainder of this paper.
Finally, it was shown in \cite{Ovrut:2012wn} that, after a coordinate transformation to a new variable $z$ with the same range $[0,\pi \rho]$ as $y$, as well as further restrictions on the coefficients, the metric can be expressed simply as
\begin{equation}
ds^{2}=a(z)^{2} \eta_{\mu\nu} dx^{\mu}dx^{\nu} + dz^{2} \ ,
\label{15}
\end{equation}
where 
\begin{equation}
a^{2}(z)=(1-2\alpha z)^{1/3} \ .
\label{16}
\end{equation}
This is the form of the five-dimensional bulk space metric that we will use in the remainder of this paper. 

Finally, this metric has two important properties that will be essential in our analysis of the heterotic three-brane worldvolume action. First, note that the only mass scale entering the metric and, hence, the curvature of the bulk space is $\alpha$--given in \eqref{11},\eqref{12}. Second,  in order to avoid metric \eqref{15} becoming singular, it follows from \eqref{16} that
\be
\alpha z < \frac{1}{2}, \quad z \in [0,\pi\rho] \ .
\label{16a}
\ee
Furthermore, as shown in \cite{Lukas:1998tt}, the ``linearized'' approximation  necessitated by the fact that $h^{1,1}=3$, strengthens this inequality to become
\be
\alpha z \ll 1 \ .
\label{16b}
\ee
For the heterotic $M$-theory standard model discussed above, we find that $1/\pi \rho \sim 10^{15}$ GeV and, using \eqref{11} and \eqref{12}, that
\be
\alpha \simeq 10^{14} \, \mathrm{GeV} \ ,
\label{BV1}
\ee
thus satisfying the inequality \eqref{16b}.


\section{Heterotic Bosonic Brane Action}
\label{sec-3}


Before constructing the $N=1$ supersymmetric worldvolume action of a probe brane in heterotic $M$-theory, we first determine the DBI action for a single real scalar degree of freedom, $\pi$, in this context. 
To do this, we utilize the formalism presented in \cite{Goon:2011qf,Goon:2011uw,deRham:2010eu} in which the DBI Galileons and the DBI conformal Galileons were constructed from probe branes in five-dimensional Minkowski space $M_{5}$ and $AdS_5$ space respectively. The heterotic DBI action, although involving nonlinear functions of $\p \pi$, will nevertheless yield two-derivative equations of motion for the scalar $\pi$. The derivatives of $\pi$ will be suppressed by the natural mass parameter $\alpha$. This will allow us to perform a derivative expansion of the heterotic DBI action in powers of $\partial/\alpha$. The resulting set of Lagrangians have a similar derivative structure to the Galileons and conformal Galileons and, hence, we will refer to them as  ``heterotic Galileons". We begin by constructing the corresponding DBI action.

\subsection{Geometric DBI Lagrangians}

We briefly review the formalism given in \cite{Goon:2011qf,Goon:2011uw,deRham:2010eu} which describes the construction of the worldvolume action of a 3-brane in a five-dimensional bulk space.
Following the conventions outlined in \cite{Deen:2017jqv}, we label the bulk space coordinates by $X^A$, $A=0,1,2,3,5$ and the brane worldvolume coordinates by $\sigma^\mu$, $\mu = 0, 1, 2, 3$. 
The bulk space is taken to be foliated by time-like hypersurfaces which are Gaussian normal with respect to the bulk metric $G_{AB}(X)$. Furthermore, we restrict to the case where the extrinsic curvature on the foliation leaves is proportional to the induced metric. The bulk space metric then takes the form
\bea
G_{AB}(X) dX^A dX^B = f(X^5)^2 g_{\mu \nu} (X) dX^{\mu} dX^{\nu }+(dX^5)^2,
\label{metric}
\eea
where $X^{\mu}$, $\mu = 0, 1, 2, 3$ are the coordinates on an arbitrary leaf of the foliation and $X^5$ is the transverse normal coordinate. The metric on the foliation, $g_{\mu\nu}(X)$, depends only on the leaf coordinates $X^\mu$ and, together with the function $f(X^5)$, will be specified once the bulk space and specific foliation are chosen. 

The worldvolume of a 3+1 brane embedded in the bulk space, parametrized by the intrinsic coordinates $\sigma^\mu$, will be labelled by five functions $X^A(\sigma)$. As a result of this embedding, the brane inherits an induced metric $\bar{g}_{\mu \nu}$ and extrinsic curvature $K_{\mu \nu}$.  Recall from \cite{Goon:2011qf,Goon:2011uw,deRham:2010eu} that if the brane action is to be invariant under worldvolume diffeomorphisms, it must consist of geometrical quantities constructed from $\bar{g}_{\mu\nu}$ and $K_{\mu \nu}$. 
Additionally, \textit{in order that the Lagrangian give rise to two-derivative equations of motion only}, the total action must be of the form \cite{deRham:2010eu}
\bea
\La = \sum_{i=1}^{5} c_i \La_i \, 
\label{CO1}
\eea
where
\bea
{\cal{L}}_1 &=& \sqrt{-g}\int^{\pi}d\pi' f(\pi')^{4},
\nn \\
{\cal{L}}_2 &=& -\sqrt{-{\bar{g}}},
\nn \\
{\cal{L}}_3 &=& \sqrt{-{\bar{g}}} ~K,
\nn \\
{\cal{L}}_4 &=& -\sqrt{-{\bar{g}}} ~ {\bar{R}},
\nn \\
{\cal{L}}_5 &=& \frac{3}{2}\sqrt{-{\bar{g}}}~{K}_{GB}
\label{La-geometric}
\eea
with $K={\bar{g}}^{\mu\nu}K_{\mu\nu}$, ${\bar{R}}={\bar{g}}^{\mu\nu}{\bar{R}}^{\alpha}_{\mu\alpha\nu}$ and ${\cal{K}}_{GB}$ is a Gauss-Bonnet boundary term given by
\be
{\cal{K}}_{GB}= -\frac{1}{3}K^{3} +K_{\mu\nu}^{2}K- \frac{2}{3} K_{\mu\nu}^{3}-2\Big( {\bar{R}}_{\mu\nu}-\frac{1}{2}{\bar{R}}{\bar{g}}_{\mu\nu} \Big) K^{\mu\nu} \ .
\ee
All indices are raised and the traces taken with respect to ${\bar{g}}^{\mu\nu}$. In this paper, we will ignore the ${\cal{L}}_{5}$ term due to its complicated structure, and examine the phenomenology of the first four Lagrangians.

Using the diffeomorphism invariance of the worldvolume action, one can set
\bea
X^{\mu} &=& \sigma^{\mu} \,,  \mu = 0, 1, 2, 3 \qquad \qquad
\nn \\
X^5 &=& \pi(\sigma) = \pi(X^\mu) \,.
\label{gauge-choice}
\eea
That is, we work in a gauge where a single scalar $\pi$, itself a function of the four foliation coordinates $X^\mu$, represents the position of the 3+1 brane with respect to the origin of the $X^A$ coordinates. Note that equation \eqref{gauge-choice} implies that $\pi$ has dimension of length; that is, mass dimension -1. It is important to note that although in a maximally symmetric bulk space, such as $AdS_{5}$, the location of the coordinate origin is completely arbitrary and carries no intrinsic information, this is no longer true in the heterotic $M$-theory bulk space. In the heterotic case, the scalar field $\pi(X^\mu)$ represents the physical location of the three-brane relative to the origin, which we choose to be located at the observable orbifold plane. For clarity, we relate our notation to that  which often appears in the literature. With this in mind, we will denote the four foliation coordinates and the transverse Gaussian normal coordinates as $X^{\mu}=x^{\mu}$, $\mu=0,1,2,3$ and $X^{5}=\pi$ respectively. It follows that the generic bulk space metric appearing in \eqref{metric} can now be written as 
\bea
G_{AB}(X) dX^A dX^B = f(\pi(x))^2 g_{\mu \nu} (x) dx^{\mu} dx^{\nu }+d\pi^2 \ .
\label{metric2}
\eea

With this gauge choice, and given a bulk space metric of the form \eqref{metric2}, the first four geometric Lagrangians are found to be \cite{Goon:2011qf, Goon:2011uw,deRham:2010eu}
\bea
{\cal{L}}_1 &=&  \sqrt{-g} \int^{\pi} d\pi' f^4(\pi') 
\nn \\
\nn \\
{\cal{L}}_2 &=& - \sqrt{-g} f^4 \sqrt{1 + \frac{1}{f^{2}} (\nabla \pi)^2}
\nn \\
\nn \\
{\cal{L}}_3 &=&  \sqrt{-g} \bigg( 
f^3 f' (5 - \gamma^2) - f^2 [\Pi] + \gamma^2 [\pi^3] \bigg)
\nn \\
\nn \\
{\cal{L}}_4 &=& -\sqrt{-g} \bigg\{ \frac{1}{\gamma}f^{2}R -2 \gamma R_{\mu\nu} \nabla^{\mu}\pi \nabla^{\nu} \pi+
\gamma \bigg(
[\Pi]^2 - [\Pi^2] + 2 \gamma^2 \frac{1}{f^{2}} \big(  - [\Pi] [\pi^3] + [\pi^4]\big)
\bigg) 
\nn \\
&&+ 6 \frac{f^3 f''}{ \gamma} (-1 + \gamma^2)+  2\gamma f f' \bigg( -4 [\Pi] + \frac{\gamma^2}{ f^{2}} \big( f^2 [\Pi] + 4 [\pi^3]\big)\bigg)
\nn \\
&&-6\frac{f^2 (f')^2}{\gamma} (1- 2\gamma^2 + \gamma^4) \bigg \} \, .
\label{La-geometric-2}
\eea
Here, $\gamma = 1/\sqrt{1 + f^{-2} (\p \pi)^2}$. The covariant derivatives and curvatures are with respect to the foliation metric $g_{\mu\nu}$.
The notation $[\Pi]$, $[\pi^n]$ is standard in the literature, see for example \cite{Goon:2011qf,deRham:2010eu}, and denote traces and contractions of derivatives of $\pi$ with respect to $g_{\mu \nu}$.  Note that the Lagrangians ${\cal{L}}_1,  {\cal{L}}_2,  {\cal{L}}_3$ and ${\cal{L}}_4$  in \eqref{La-geometric-2} have mass dimensions $-1, 0, 1$ and $2$ respectively. Hence, the constant coefficients  $c_{1}, c_{2},c_{3}, c_{4}$ in the action \eqref{CO1} have mass dimensions $5, 4, 3$ and $2$. 

The formalism and results described thus far are valid for a probe three-brane in any background five-dimensional bulk space. We now apply this generic formalism to the case of a probe three-brane embedded in the five-dimensional bulk space of heterotic $M$-theory. It follows from the metric \eqref{15}, \eqref{16} presented in Section 2 that 
\be
f(\pi) = (1-2\alpha \pi)^{1/6}
\label{C1}
\ee
and, hence,
\be
\gamma = \frac{1}{\sqrt{1 + (1- 2\alpha \pi)^{-1/3} (\p \pi)^2}} \ .
\label{C2}
\ee
The total Lagrangian then becomes
\be
{\cal{L}}=\sum_{i=1}^{4} {\cal{L}}_{i}
\label{CO2}
\ee
where the geometric Lagrangians presented in \eqref{La-geometric-2} are given by
\bea
\La_1 &=& 
-\frac{3}{10 \alpha } (1 - 2\alpha \pi )^{5/3}
\nn \\
\nn \\
\La_2 &=& 
- (1 - 2 \alpha \pi )^{2/3} \sqrt{1 + (1- 2\alpha \pi)^{-1/3} (\p \pi)^2}
\nn \\
\nn \\
\La_3 &=&
-\frac{\alpha }{3} (1 - 2 \alpha \pi )^{-1/3} \big[ 5 - \gamma^2 \big]
- (1 - 2 \alpha \pi)^{1/3} \Box \pi 
+ \gamma^2 [\pi^3]
\nn \\
\La_4 &=&
- \gamma
\bigg(
 [\Pi]^2 - [\Pi^2] 
+ 2 \gamma^2 (1- 2\alpha \pi)^{-1/3} \big[ - [\Pi] [\pi^3] +  [\pi^4]\big]
\bigg)
\nn \\
&&
+ \frac{10}{3}\frac{\alpha^2}{\gamma} (1 - 2 \alpha \pi)^{-4/3} (-1 + \gamma^2)
\nn \\
&&
+ \frac{2}{3}\alpha \gamma (1 - 2 \alpha \pi)^{-2/3} \bigg( -4 \Box \pi + \gamma^2 \big[\Box \pi + 4  (1 - 2 \alpha \pi)^{-1/3} [\pi^3] \big]\bigg)
\nn \\
&&
+ \frac{2}{3} \frac{\alpha}{\gamma} (1 - 2 \alpha \pi )^{-4/3}\big( 1 - 2\gamma^2 + \gamma^4 \big) \, .
\label{DBI-heterotic}
\eea
Note that, unlike the Poincare and conformal DBI Galileons, these Lagrangians do not exhibit a non-linearly realized global symmetry. The reason is that the Poincare and conformal symmetries arise from those Killing vectors of the bulk space which are not parallel to the surfaces of foliation. While such Killing vectors are present in the maximally symmetric $M_{5}$ and $AdS_5$ spaces, there are none in the heterotic bulk space. Hence, the absence of an analogous symmetry in the heterotic DBI Galileons. We emphasize that since the Lagrange densities in \eqref{DBI-heterotic} arise from those presented in \eqref{La-geometric}, the associated equations of motion all contain at most two derivatives.

\subsection{The Derivative Expansion}

In a previous paper \cite{Deen:2017jqv}, we performed a derivative expansion of the DBI Galileon action in the case of a probe three-brane embedded in an $AdS_{5}$ bulk space. There, one takes all derivatives of the brane modulus field to be much smaller than the mass scale $\mathcal{M} = 1/{\mathcal{R}}$, where $\mathcal{R}$ is the radius of curvature of the $AdS_5$ space. The total worldvolume DBI Lagrangian was expanded in powers of $(\p/\mathcal{M} )^2$, and terms of the same order in derivatives were then grouped together. It was shown that, at each order $n$ in derivatives, the associated Lagrangian was precisely the $n$-th order conformal Galileon. Due to the symmetry properties of the complete worldvolume action, one need only expand to order $(\p / \mathcal{M})^8$--corresponding to the conformal Galileon ${\cal{L}}_{5}$--since all terms of higher order form a total divergence \cite{Deffayet:2009wt}.

For the case of heterotic M-theory, the mass scale associated with the curvature of the five-dimensional bulk space is $\alpha$, as discussed in Section 2. Hence, the appropriate expansion parameter in the heterotic case will be $(\p/\alpha)^2$. As discussed above, there is no special symmetry inherent in heterotic geometry. One might think, therefore, that a derivative expansion of the Lagrangians in \eqref{DBI-heterotic} would require one to keep terms to all order in $(\p /\alpha)^2$. 
However, this is not the case. Heterotic $M$-theory is only valid for momenta that are small compared to, not only the Planck mass $M_{P}$ and the Calabi-Yau scale of order $10^{16}$ GeV, but also with respect to the scale associated with the curvature of the fifth-dimension. As discussed above, for the heterotic standard model this is found to be of order $10^{14}$ GeV. 
Therefore, it is necessary to restrict $(\p /\alpha)^2$ to be small and, hence, one can truncate the derivative expansion at a small finite order in this expansion parameter.

We begin by defining the dimensionless field
\bea
\hpi = \alpha \pi \, .
\eea
Let us also scale the individual Lagrangians $\La_i$ and coefficients $c_i$ as follows
\bea
\La_i  \rightarrow \alpha^{2-i} \La_i \, , \qquad
c_i \rightarrow \alpha^{i-2} c_i   
\eea
for $i=1,2,3,4$. This ensures that the $c_i$, while still arbitrary, now have mass dimension 4, while each Lagrangian density $\La_i$ is dimensionless.
We now expand the total Lagrangian $\La$ in powers of 
\be
(\p /\alpha )^2 \ll 1. 
\label{noise1}
\ee
Collecting terms up to order $( \p /\alpha)^6$, and using integration by parts, we can then express our total Lagrangian \eqref{CO2} as
\bea
\La &=& \sum_{i=1}^4 \overbar{\La}_{i}
\eea
where
\bea
\overbar{\La}_{1} &=&
-\frac{3}{10} c_1(1 - 2\hpi)^{5/3} - c_2 (1 - 2\hpi)^{2/3} - \frac{4}{3} c_3 (1- 2 \hpi)^{-1/3} 
\nn \\
\nn \\
\overbar{\La}_{2} &=&
 \bigg(
 - \frac{1}{2}c_2(1 - 2\hpi)^{1/3} - c_3 (1- 2\hpi )^{-2/3} - \frac{2}{3} c_4 (1 - 2\hpi)^{-5/3} 
\bigg)\bigg(\frac{\p \hpi}{\alpha}\bigg)^2
\nn \\
\nn \\
\overbar{\La}_{3} &=&
\bigg( - \frac{1}{2}c_3 - c_4 (1 - 2 \hpi)^{-1} \bigg)  \bigg(\frac{\p \hpi}{\alpha} \bigg)^2 \frac{\square \hpi}{\alpha^2}
+
\bigg( \frac{1}{8}c_2  + \frac{1}{3} c_3 (1 - 2\hpi)^{-1} - \frac{1}{3} c_4  (1 - 2\hpi)^{-2} \bigg) \bigg(\frac{\p \hpi}{\alpha} \bigg)^4 
\nn \\
\nn 
\eea
\bea
\overbar{\La}_{4} &=&
- \frac{1}{4} c_4 (1- 2\hpi)^{-1/3} \frac{\p_\mu}{\alpha} \bigg(\frac{\p \hpi}{\alpha} \bigg)^2 \frac{\p^\mu}{\alpha} \bigg(\frac{\p \hpi}{\alpha} \bigg)^2
+ c_4 (1- 2\hpi)^{-1/3} \frac{\Box \hpi}{\alpha^2} \frac{\hpi^{, \mu}}{\alpha} \frac{\hpi_{, \mu \nu} }{\alpha^2} \frac{\hpi^{, \nu}}{\alpha} \qquad \qquad \qquad
\nn \\
&-&
\frac{19}{6}c_4 (1- 2\hpi)^{-4/3} \bigg(\frac{\p \hpi}{\alpha} \bigg)^4 \frac{\Box \hpi}{\alpha^2}
\nn \\
& +&
\bigg( - c_3 (1- 2\hpi)^{-1/3} - \frac{11}{3}c_4 (1- 2\hpi)^{-4/3} \bigg) \bigg(\frac{\p \hpi}{\alpha} \bigg)^2 \frac{\hpi^{, \mu}}{\alpha} \frac{\hpi_{,\mu \nu}}{\alpha^2} \frac{\hpi^{, \nu}}{\alpha}
\nn \\
&+& 
\bigg( - \frac{1}{16} c_2 (1- 2\hpi)^{-1/3} - \frac{1}{3} (1- 2\hpi)^{-4/3} - \frac{9}{4} c_4 (1- 2\hpi)^{-7/3} \bigg) \bigg(\frac{\p \hpi}{\alpha} \bigg)^6 \, .
\label{La-derivative}
\eea

Up to now, we have discussed the derivative expansion of the DBI heterotic Lagrangian using the necessary restriction that $(\partial/\alpha)^{2} \ll 1$. However, there is an additional physical restriction that must be taken into account. It follows from \eqref{16b} that the dimensionless field $\hpi$ must satisfy
\be
\hpi \ll 1 \ .
\label{VT2}
\ee
While the DBI expressions given in \eqref{DBI-heterotic} can be considered accurate as far as the expansion in $(\p / \alpha)^{2}$ is concerned,  we must now additionally expand all functions of $\hpi$ derived from $f(\hpi)$ and its derivatives to {\it linear order} in $\hpi$. This expansion must terminate at linear order since higher powers
of  $\hpi$ cannot arise in the metric deduced from the dimensional reduction of $M$-theory to leading order in $\kappa$. Performing this expansion in \eqref{La-derivative}, we find that the worldvolume Lagrangian of a probe three-brane in five-dimensional heterotic $M$-theory is given by 
\bea
\La = \sum_{i=1}^4 \bar{\La}_i \, 
\eea
where
\bea
\bar{\La}_{1} &=& 
- \frac{3}{10} c_1 - c_2 - \frac{4}{3}c_3 
+ \big( c_1 + \frac{4}{3} c_2 - \frac{8}{9} c_3 \big) \hpi
\nn \\
\nn \\
\bar{\La}_{2} &=& 
\bigg(- \frac{1}{2} c_2 -c_3 - \frac{2}{3}c_4 + \big( \frac{1}{3} c_2 - \frac{4}{3} c_3 - \frac{20}{9} c_4 \big) \hpi \bigg) \bigg(\frac{\p \hpi}{\alpha}\bigg)^2
\nn \\
\nn \\
\bar{\La}_{3} &=&
\bigg(-\frac{1}{2}c_3 -  c_4 -2 c_4 \hpi \bigg)\bigg(\frac{\p \hpi}{\alpha}\bigg)^2 \frac{\square \hpi}{\alpha^2}
+ \bigg( \frac{1}{8} c_2 + \frac{1}{3}c_3 -  \frac{1}{3}c_4 + (\frac{2}{3}c_3 -  \frac{4}{3}c_4) \hpi \bigg) \bigg(\frac{\p \hpi}{\alpha}\bigg)^4 
\nn \\
\nn \\
\bar{\La}_{4} &=&
- \bigg(\frac{1}{4} c_4 + \frac{1}{6} c_4 \hpi \bigg) \frac{\p_\nu}{\alpha} \bigg(\frac{\p \hpi}{\alpha}\bigg)^2 \frac{\p^\nu}{\alpha} \bigg(\frac{\p \hpi}{\alpha}\bigg)^2 
+ \bigg(c_4 + \frac{2}{3} c_4 \hpi \bigg) \frac{\Box \hpi}{\alpha^2} \frac{\hpi^{, \mu}}{\alpha}\frac{ \hpi_{, \mu \nu}}{\alpha^2} \frac{\hpi^{, \nu}}{\alpha}
\nn \\
&&
- \bigg(\frac{19}{6} c_4 + \frac{76}{9} c_4 \hpi \bigg) \bigg(\frac{\p \hpi}{\alpha}\bigg)^4 \frac{\Box \hpi}{\alpha^2}
+ \bigg(- c_3 - \frac{11}{3} c_4 + (-\frac{2}{3} c_3 - \frac{88}{9} c_4) \hpi \bigg) \bigg(\frac{\p \hpi}{\alpha}\bigg)^2 \frac{\hpi^{, \mu }}{\alpha} \frac{\hpi_{,\mu \nu}}{\alpha^2} \frac{\hpi^{, \nu}}{\alpha}
\nn \\
&&
+\bigg(-\frac{1}{16} c_2 - \frac{1}{3} c_3 - \frac{9}{4} c_4  + (-\frac{1}{24} c_2 - \frac{8}{9} c_3 - \frac{21}{2}c_4) \hpi \bigg)
\bigg(\frac{\p \hpi}{\alpha}\bigg)^6 \ .
\nn \\
\label{heterotic-galileons}
\eea
Again, we note the absence of a non-linearly realized global symmetry in the worldvolume Lagrangian. In the small derivative limit, this means that, unlike in the conformal case, one cannot re-express the $c_i$ coefficients in terms of new constants $\bar{c}_i$ such that the total Lagrangian is of the form $\sum_{i=1}^{4} \bar{c}_i \bar{\La}_i$. This feature of the parameters will be helpful when the formalism is used in a cosmological context--such as to ensure the appearance of NEC violation. Be that as it may, since the expressions in \eqref{heterotic-galileons} arise from those in \eqref{DBI-heterotic}, they give rise to second order equations of motion and make up the worldvolume action for a probe brane in a five-dimensional heterotic $M$-theory geometry. Therefore, we will refer to  them as ``heterotic Galileons".


\section{Supersymmetric Heterotic Galileons}

We now extend the scalar Lagrangians given in \eqref{heterotic-galileons} to d=4,  $N=1$ global supersymmetry, as is required by heterotic $M$-theory. To do this, we employ the formalism of $N=1$ superspace \cite{Wess}, whose coordinates are $x^{\mu}$, $\mu=0,1,2,3$, an anti-commuting two component Weyl spinor $\theta$ and its hermitian conjugate $\bar{\theta}$. These coordinates have mass dimensions 0, -1/2 and 1/2 respectively. Following \cite{ Deen:2017jqv, Khoury:2011da}, we begin by defining a dimensionless complex scalar field $A$, whose real part is the brane position modulus $\hpi$. That is,
\bea
A(x) \equiv \frac{1}{\sqrt{2}} (\hpi(x) + i \chi (x) ) \ ,
\eea
where $\chi$ is a second real scalar field. We now take the scalar field $A$ to be the lowest component of a dimensionless chiral superfield $\Phi(x^{\mu},\theta,\bar{\theta})$. Expressing this as an expansion in the anti-commuting spinor coordinates, one finds that there are two new fields in the chiral multiplet in addition to $A$. These are a complex two-component Weyl spinor $\psi$ and a complex scalar field $F$, with mass dimensions 1/2 and 1 respectively. Abusing notation, we can simply write
\bea
\Phi = (A, \psi, F) \, .
\eea

Using the superspace formalism, one can construct manifestly supersymmetric Lagrangians as the $\theta \theta \bar{\theta} \bar{\theta}$ component of a {\it real combination} of $\Phi$ and $\Phi^\dagger$ (such as $\Phi \Phi^\dagger$), or the $\theta \theta$ (or $\bar{\theta }\bar{\theta}$) component of a {\it complex, holomorphic} function of $\Phi$ (or $\Phi^\dagger$) alone (such as $\Phi^{2})$. Henceforth, since it is not required in this paper, we will drop all terms involving the fermion $\psi$ and focus on the bosonic fields only.

\subsection{Supersymmetric ${\bar{{\cal{L}}}}_{2}$}

We start by defining a manifestly hermitian K\"ahler potential by
\bea
K(\Phi, \Phi^\dagger) &=& 
\frac{(c_2 + 2 c_3 +\frac{4}{3}c_{4})}{\alpha^2} \Phi \Phi^\dagger 
+ \frac{1}{\sqrt{2}} \frac{(- \frac{1}{3}c_2 + \frac{4}{3}c_3 + \frac{20}{9} c_4)}{\alpha^2} (\Phi^2 \Phi^\dagger + \Phi \Phi^{\dagger 2}) \, .
\label{p1}
\eea
Note that $K$ is a real superfield. The supersymmetric extension of the $\bar{\La}_2$ Lagrangian in \eqref{heterotic-galileons} is then the highest (that is, $\theta \theta \bar{\theta} \bar{\theta}$) component of $K(\Phi, \Phi^\dagger)$, given by
\bea
\overbar{\La}_2^{\rm{SUSY}} &=& K(\Phi, \Phi^\dagger) \bigg|_{\theta \theta \bar{\theta} \bar{\theta}} 
= - \frac{\p^2 K}{\p A \p A^*} \p A \cdot \p A^* + \frac{\p^2 K}{\p A \p A^*} FF^* 
\nn \\
&=&
\bigg(- \frac{1}{2} c_2 -  c_3  - \frac{2}{3}c_4+ ( \frac{1}{3}c_2 - \frac{4}{3}c_3 - \frac{20}{9} c_4) \hpi \bigg)
\bigg( 
 (\frac{\p \hpi}{\alpha})^2 + (\frac{\p \chi}{\alpha})^2 - 2\frac{F F^*}{\alpha^2}
\bigg) \ .
\label{L2-susy}
\eea
It is important to note that the dimension one auxiliary field $F$ that appears here is everywhere suppressed by $\alpha$, in the same manner as the derivatives $\partial\hpi$. 
To simplify the notation, for the remainder of this section, unless explicitly stated otherwise, we will set $\alpha =1$.

\subsection{Supersymmetric ${\bar{{\cal L}}}_3$}

The supersymmetric extension of the $\bar{\La}_3$ Lagrangian given in \eqref{heterotic-galileons} can be constructed from two terms, 
\bea
\overbar{\La}_{3, \mathrm{1st\, term}}^{\rm{SUSY}} &=&
 \frac{1}{16}
\bigg( - \frac{1}{\sqrt{2}}c_3 - \frac{ \sqrt{2}}{3} c_4 -2 c_4 (\Phi + \Phi^\dagger)\bigg)
\bigg( D \Phi D \Phi \overbar{D}^2 \Phi^{\dagger} + \mathrm{h.c.}\bigg)\bigg|_{\theta \theta \bar{\theta} \bar{\theta}} 
\nn \\
&=&
\bigg(- \frac{1}{\sqrt{2}}c_3 - \frac{ \sqrt{2}}{3} c_4 -2  c_4 (A + A^*)\bigg) \bigg(
(\p A)^2 \square A^* + (\p A^*)^2 \square A  
- F F^* (\square A  +  \square A^* )
\nn \\ 
&&
\qquad \qquad \qquad 
+ F^* \p F \cdot (\p A -   \p A^*) 
- F \p F^* \cdot(  \p A -\p A^* )
\bigg)
\nn \\
&+& 4 c_4 
\bigg( 
 (FF^*)^2 - FF^* \p A \cdot \p A^*
\bigg)
\eea
and
\bea
\overbar{\La}_{3, \mathrm{2nd\, term}}^{\rm{SUSY}} &=&
 \frac{1}{4} 
\bigg( \frac{1}{8} c_2 + \frac{1}{3}c_3 - \frac{1}{3} c_4 + \frac{1}{\sqrt{2}} \big(\frac{2}{3}c_3 - \frac{4}{3}c_4 \big) 
 (\Phi + \Phi^\dagger) \bigg)
 \bigg( D \Phi D \Phi \bar{D} \Phi^\dagger \bar{D} \Phi^\dagger \bigg)\bigg|_{\theta \theta \bar{\theta} \bar{\theta}}
\nn \\
 &=&
 \bigg( \frac{1}{8} c_2 + \frac{1}{3}c_3 - \frac{1}{3} c_4 + \frac{1}{\sqrt{2}} \big(\frac{2}{3}c_3 - \frac{4}{3}c_4 \big)   (A + A^*) \bigg)
 \bigg( 4(FF^*)^2 - 8 FF^* \p A \cdot \p A^* + 4 (\p A)^2 (\p A^*)^2\bigg) \
\nn \\
\eea
In terms of the real scalar fields $\hpi$ and $\chi$, as well as the complex auxiliary field $F$, we find that the complete supersymmetrization of $\bar{\La}_3$ is given by
\bea
\overbar{\La}_{3}^{\rm{SUSY}} &=&
\bigg( - \frac{1}{2}c_3 - c_4 - 2 c_4 \hpi \bigg)
 \bigg(
(\p \hpi)^2 \Box \hpi + (\p \chi)^2 \Box \hpi + 2 \p \hpi \cdot \p \chi \Box \chi
- 2F F^* \Box \hpi
\nn \\
&&
\qquad \qquad \qquad 
+ 2iF^* \p F \cdot \p \chi 
-  2i F \p F^* \cdot \p \chi
\bigg)
\nn \\
&+&4 c_4 
\bigg( 
 (FF^*)^2 + \frac{1}{2} FF^* (\p \hpi)^2 + \frac{1}{2}FF^* (\p \chi)^2  
\bigg)
\nn \\
&+&  \bigg( \frac{1}{8} c_2 + \frac{1}{3}c_3 - \frac{1}{3} c_4 + \big(\frac{2}{3}c_3 - \frac{4}{3}c_4 \big) \hpi \bigg)
\bigg((\p \hpi)^4 + (\p \chi)^4 - 2 (\p \hpi )^2 (\p \chi)^2 + 4(\p \hpi \cdot \p \chi)^2  
\nn \\
&&
\qquad \qquad \qquad 
 -   4 FF^* (\p \hpi)^2 - 4 FF^* (\p \chi)^2    +  4 (FF^*)^2 \bigg) \, .
\label{L3-susy}
\eea
We note the appearance of derivatives of $F$, as well as a term proportional to $(FF^*)^2$. In the conformal Galileon case arising from the $AdS_{5}$ bulk space, the first type of term occurred at the level of $\overbar{\La}_{3}^{\rm{SUSY}} $, but a serendipitous cancellation removed the latter type at this order.

\subsection{Supersymmetric ${\overbar{\La}}_4$}

In order to supersymmetrize the fourth order heterotic Galileon, we have to consider each of the five terms in \eqref{heterotic-galileons} that comprise $\bar{\La}_4$ separately.
Let us begin with the term involving $\p_\mu (\p \hpi)^2 \p^\mu (\p \hpi)^2$. To extend this to $N=1$ supersymmetry, we construct
\begin{eqnarray}
\overbar{\La}_{4, \, \mathrm{1st\,term}}^{\mathrm{SUSY}}
&=& 
\frac{1}{32} \bigg( \frac{1}{4} c_4 + \frac{1}{6\sqrt{2}} c_4 (\Phi + \Phi)\bigg)
\{ D, \bar{D}\} (D \Phi D \Phi ) \{ D, \bar{D} \} (\bar{D} \Phi^\dagger \bar{D} \Phi^\dagger)\bigg|_{\theta\theta\bar{\theta}\bar{\theta}}
\nn \\
&=&
\bigg( - \frac{1}{4} c_4 - \frac{1}{6\sqrt{2}} c_4 (A + A^*)\bigg)
\bigg(
4\p_\mu (\p A)^2 \p^\mu (\p A^*)^2 
-8 \p_\mu (F A_{,\nu}) \p^\mu (F^* A^{*, \nu})
+ 16 F F^* \p F \cdot \p F^*
\bigg)
\nn \\
\nn \\
&=& 
\frac{1}{32} \bigg( \frac{1}{4} c_4 + \frac{1}{6\sqrt{2}} c_4 (\Phi + \Phi)\bigg)
\{ D, \bar{D}\} (D \Phi D \Phi ) \{ D, \bar{D} \} (\bar{D} \Phi^\dagger \bar{D} \Phi^\dagger)\bigg|_{\theta\theta\bar{\theta}\bar{\theta}}
\nn \\
&=&
\bigg( - \frac{1}{4} c_4 - \frac{1}{6\sqrt{2}} c_4 (A + A^*)\bigg)
\bigg(
4\p_\mu (\p A)^2 \p^\mu (\p A^*)^2 
-8 \p_\mu (F A_{,\nu}) \p^\mu (F^* A^{*, \nu})
+ 16 F F^* \p F \cdot \p F^*
\bigg)
\nn \\
\nn \\
&=&
\bigg( - \frac{1}{4} c_4 - \frac{1}{6} c_4  \hpi \bigg)
\bigg(
\p_\mu (\p \hpi)^2 \p^\mu (\p \hpi)^2 + \p_\mu (\p \chi)^2 \p^\mu (\p \chi)^2 - 2 \p_\mu (\p \hpi)^2 \p^\mu (\p \chi)^2 
\nn \\
&&
\qquad  \qquad \qquad \quad
+ 4 \p_\mu (\p \hpi \cdot \p \chi )\p^\mu (\p \hpi \cdot \p \chi) 
- 4 \p_\mu (F \hpi_{,\nu}) \p^\mu (F^* \hpi^{,\nu})
\nn \\
&&
\qquad \qquad \qquad \quad
- 4 \p_\mu (F \chi_{,\nu}) \p^\mu (F^* \chi^{,\nu})
+ 16 F F^* \p F \cdot \p F^*
\bigg) \ .
\label{L4-1}
\end{eqnarray}
In addition to the desired term $\p_\mu (\p \hpi)^2 \p^\mu (\p \hpi)^2$, as well as related terms containing both scalars $\hpi$ and $\chi$, we encounter terms involving two derivatives of $F$ in this expression; for example, $F F^* \p F \cdot \p F^*$. These will occur throughout the supersymmetrization of $\bar{\La}_4$.

Next, we consider the term involving $\Box \hpi \p^\mu (\p \hpi)^2 \hpi_{, \mu}$. It is given by
\bea
\nn
{\overbar{\La}}_{4,\,{\rm 2nd}\;{\rm term}}^{\rm SUSY}  &=& 
\frac{1}{128} \bigg( c_4 + \frac{2}{3\sqrt{2}} c_4 (\Phi + \Phi^\dagger) \bigg)
\bigg( \{D, \bar{D} \} (\Phi + \Phi^\dagger)  \{D, \bar{D} \} (D \Phi D \Phi ) \bar{D}^2 \Phi^\dagger 
+ \mathrm{h.c.}
\bigg)
\bigg|_{\theta \theta \bar{\theta} \bar{\theta}}
\nn \\
&=&
 \bigg( c_4 +  \frac{2}{3\sqrt{2}} c_4 (A + A^*) \bigg)
\bigg(
(A + A^*)^{, \mu} \big( \p_\mu (\p A)^2 \Box A +  \p_\mu (\p A^*)^2 \Box A^*\big)
\nn \\
&&
\qquad \qquad \qquad 
- (A + A^*)^{, \mu} \big( \p_\mu (F A_{, \nu}) F^{*, \nu} + \p_\mu (F^* A_{, \nu}^*) F^{, \nu}  \big)
\nn \\
&&
\qquad \qquad \qquad 
-  (A + A^*)^{, \mu} \big( \p_\mu (F \Box A - \p F \cdot \p A) F^*
+ \p_\mu (F^* \Box A^* - \p F^* \cdot \p A^*) F \big)
\nn \\
&&
\qquad \qquad \qquad 
- \p_\mu \p_\nu (A - A^*) \big( \p^\mu (F A^{, \nu}) F^* - \p^\mu (F^* A^{*, \nu}) F \big)
- 4 F F^* \p F \cdot \p F^*
\bigg)
\nn \\
&-&\frac{32}{3\sqrt{2}} c_4 (A - A^*)^{, \nu} (A + A^*)_{, \mu} 
\big( \p^\mu (F A_\nu) F^* -  \p^\mu (F^* A_\nu^*) F \big)
\nn \\
&-& \frac{32}{3\sqrt{2}} c_4 F F^*  (A + A^*)_{, \mu}
\big(  \p^\mu (\p A)^2 + \p^\mu (\p A^*)^2  \big)
\nn \\
&+& \frac{64}{3\sqrt{2}} c_4 \bigg( F (F^*)^2 (A + A^*)_{, \mu} F^{, \mu}  +  F^* F^2 (A + A^*)_{, \mu} F^{*, \mu}\bigg)
\eea
Expressed in terms of the fields $\hpi$ and $\chi$, as well as the auxiliary field $F$, this becomes
\bea
{\overbar{\La}}_{4,\,{\rm 2nd}\;{\rm term}}^{\rm SUSY}  &=& 
\bigg( c_4 +  \frac{2}{3} c_4 \hpi \bigg)
\bigg(
\Box \hpi \p^\mu (\p \hpi)^2 \hpi_{, \mu} - \Box \hpi \p^{\mu}(\p \chi)^2 \hpi_{, \mu} - 2\Box \chi \p^\mu (\p \hpi \cdot \p \chi) \hpi_{, \mu}
\nn \\
&&
\qquad 
- \hpi^{, \mu } \p_{\mu} (F \hpi_{, \nu}) F^{*, \nu} - \hpi^{, \mu } \p_{\mu} (F^* \hpi_{, \nu}) F^{, \nu}
- i \hpi^{, \mu } \p_{\mu} (F \chi_{, \nu}) F^{*, \nu} + i \hpi^{, \mu } \p_{\mu} (F^* \chi_{, \nu}) F^{, \nu}
\nn \\
&&
\qquad 
- \hpi^{, \mu } \p_\mu ( F \Box \hpi)F^* - \hpi^{, \mu} \p_{\mu} (F^* \Box \hpi)F
- \hpi^{, \mu } \p_\mu ( \p F \cdot \p \hpi) F^* - \hpi^{, \mu} \p_{\mu} (\p F^* \cdot \p \hpi)F
\nn \\
&&
\qquad 
- i \hpi^{, \mu } \p_\mu ( F \Box \chi)F^* + i \hpi^{, \mu} \p_{\mu} (F^* \Box \chi)F
+ i\hpi^{, \mu } \p_\mu ( \p F \cdot \p \chi) F^* - i \hpi^{, \mu} \p_{\mu} (\p F^* \cdot \p \chi)F
\nn \\
&&
\qquad 
- i \chi_{, \mu \nu} \p^{\mu}(F \hpi^{, \nu})F^* + i  \chi_{, \mu \nu} \p^{\mu}(F^* \hpi^{, \nu})F
+ \chi_{, \mu \nu} \p^{\mu}(F \chi^{, \nu})F^* +   \chi_{, \mu \nu} \p^{\mu}(F^* \chi^{, \nu})F
\nn \\
&&
\qquad 
- 4 F F^* \p F \cdot \p F^*
\bigg)
\nn \\
&+& \frac{32}{3} c_4 \bigg(
i \hpi^{, \mu} \p_{\mu} (F \hpi_{, \nu}) \chi^{, \nu} F^* - i \hpi^{, \mu} \p_{\mu} (F^* \hpi_{, \nu}) \chi^{, \nu} F
-  \hpi^{, \mu} \p_{\mu} (F \chi_{, \nu}) \chi^{, \nu} F^* - \hpi^{, \mu} \p_{\mu} (F^* \chi_{, \nu}) \chi^{, \nu} F
\bigg)
\nn \\
&-& \frac{32}{3} c_4 \bigg(
F F^* \hpi^{, \mu} \p_{\mu} (\p \hpi)^2 + F F^* \hpi^{, \mu} \p_\mu (\p \chi)^2 
\bigg)
+\frac{64}{3} c_4 \bigg(
F (F^*)2 \p F \cdot \p \hpi + F^* F^2 \p F^* \cdot \p \hpi
\bigg)
\nn \\
\label{L4-2}
\eea
The remaining three terms, which involve $ (\p \hpi)^4 \Box \hpi$, $(\p \hpi)^2 \hpi_{,\mu} \hpi_{\mu \nu} \hpi^{, \nu}$ and $(\p \hpi)^4 $respectively, have the following supersymmetric extensions.
\bea
\overbar{ \La}_{4,\,{\rm 3rd ~term}}^{\rm SUSY} &=& 
\frac{1}{32 \sqrt{2}} \bigg( \frac{19}{6} c_4 + \frac{76}{9\sqrt{2}} c_4 (\Phi + \Phi^\dagger)\bigg)
D \Phi D \Phi \bar{D} \Phi^\dagger \bar{D} \Phi^\dagger \{D, \bar{D} \}  \{D, \bar{D} \} (\Phi + \Phi^\dagger) 
\bigg|_{\theta \theta \bar{\theta} \bar{\theta}}
\nn \\
&=&
2\sqrt{2} \bigg( -\frac{19}{6} c_4 -\frac{76}{9\sqrt{2}}c_4 ( A + A^*)\bigg)
\Box (A + A^*) \bigg( (\p A )^2(\p A^*)^2 - 2 F F^* \p A \cdot \p A^* + (F F^*)^2\bigg)
\nn \\
\nn \\
&=&
 \bigg( - \frac{19}{6} c_4 - \frac{76}{9}c_4 \hpi \bigg) 
 \bigg(
 (\p \hpi)^4 \Box \hpi + (\p \chi)^4 \Box \hpi - 2 (\p \hpi)^2 (\p \chi)^2 \Box \hpi
 + (\p \hpi \cdot \p \chi)^2 \Box \hpi 
 \nn \\
 &&
 \qquad \qquad \qquad 
 - 2 F F^* ( (\p \hpi)^2 + (\p \chi)^2 ) \Box \hpi
 + 4 (F F^* )^2 \Box \hpi
 \bigg)\, ,
\label{L4-3}
\eea
\bea
\overbar{\La}_{4,\,{\rm 4th ~term}}^{\rm SUSY} &=& 
-\frac{1}{128 \sqrt{2}}
\bigg( -c_3 - \frac{11}{3}c_4 +\frac{1}{\sqrt{2}}  (-\frac{2}{3} c_3 - \frac{88}{9} c_4) (\Phi + \Phi^\dagger)
\bigg)
\nn \\
&&
\bigg(\{ D, \bar{D} \} D \Phi D \Phi \bar{D} \Phi \bar{D} \Phi^\dagger \{ D, \bar{D} \} \Phi 
+ \mathrm{h.c.}
\bigg)
\bigg|_{\theta \theta \bar{\theta} \bar{\theta}}
\nn \\
&=&
\frac{1}{\sqrt{2}} \bigg( -c_3 - \frac{11}{3} c_4 + \frac{1}{\sqrt{2}}(- \frac{2}{3} c_3 - \frac{88}{9}c_4) (A + A^*) \bigg)
\nn \\
&&
\bigg( 
\p_\mu \big( (\p A)^2 (\p A^*)^2 \big) 
-2 \p_\mu (F F^* \p A \cdot \p A^*) 
+ \p_\mu (F F^*)^2
\bigg)\big( A + A^*\big)^{, \mu}
\nn \\
\nn \\
&=&
\bigg( -c_3 - \frac{11}{3} c_4 + (- \frac{2}{3} c_3 - \frac{88}{9}c_4) \hpi  \bigg)
\bigg(
(\p \hpi)^2 \hpi_{,\mu} \hpi_{\mu \nu} \hpi^{, \nu}
- (\p \hpi)^2 \hpi_{,\mu} \hpi_{\mu \nu} \hpi^{, \nu}
\nn \\ 
&&
- (\p \hpi)^2 \hpi_{,\mu} \chi_{\mu \nu} \chi^{, \nu}
+ (\p \chi)^2 \hpi_{,\mu} \chi_{\mu \nu} \chi^{, \nu}
+ 4 \p_\mu \big( \p \hpi \cdot \p \chi \big)^2 \hpi^{, \mu}
- \p_\mu (F F^* (\p \hpi)^2 ) \hpi^{, \mu} 
\nn \\ 
&&
- \p_\mu (F F^* (\p \chi)^2 ) \hpi^{, \mu} 
+ \p_\mu(F F^*) \hpi^{, \mu}
\bigg) \, ,
\label{L4-4}
\eea
\bea
\bar{{\cal L}}_{4,\,{\rm 5th ~term}}^{\rm SUSY} 
&=& 
\frac{1}{16} \bigg( -\frac{1}{16} c_2 - \frac{1}{3} c_3 - \frac{9}{4} c_4 
+ \frac{1}{\sqrt{2}} (-\frac{1}{24} c_2 -\frac{8}{9}c_3 - \frac{21}{2} c_4) (\Phi + \Phi^\dagger)\bigg)
D \Phi D \Phi \bar{D} \Phi^\dagger \bar{D} \Phi^\dagger 
\nn \\
&&
\{ D, \bar{D} \} \Phi \{ D, \bar{D} \} \Phi^\dagger 
\bigg|_{\theta \theta \bar{\theta} \bar{\theta}}
\nn \\
&=&
\bigg( 
-\frac{1}{16} c_2 - \frac{1}{3} c_3 - \frac{9}{4} c_4 
+ \frac{1}{\sqrt{2}} (-\frac{1}{24} c_2 -\frac{8}{9}c_3 - \frac{21}{2} c_4) 
(A + A^*)
\bigg)
\nn \\
&&
\quad
\bigg(
(\p A )^2(\p A^*)^2 - 2 F F^* \p A \cdot \p A^* + (F F^*)^2
\bigg) \p A \cdot \p A^*
\nn \\
\nn \\
&=&
\bigg(
-\frac{1}{16} c_2 - \frac{1}{3} c_3 - \frac{9}{4} c_4 
+ (-\frac{1}{24} c_2 -\frac{8}{9}c_3 - \frac{21}{2} c_4)  \hpi
\bigg)
\bigg(
(\p \hpi)^2 + (\p \chi)^2
\bigg)
\nn \\
&&
\, \, \, 
\bigg( 
(\p \hpi)^4 + (\p \chi)^4 - 2 (\p \hpi)^2 (\p \chi)^2 + 4 (\p \hpi \cdot \p \chi)^2 
- 4 F F^* \big( (\p \hpi)^2 - (\p \chi)^2\big)
+ 4 (F F^*)^2
\bigg) \, .
\nn \\ 
\label{L4-5}
\eea

\subsection{Supersymmetric ${\bar{\cal{L}}}_{1}$}

Thus far, we have ignored the first scalar Lagrangian density  ${\bar{\cal{L}}}_{1}$ given in \eqref{heterotic-galileons}. Since ${\bar{\cal{L}}}_{1}$ is a function of $\hpi$ only, without any derivatives, it is logical to treat it as a potential energy term for $\hpi$. In $N=1$ supersymmetry, one specifies a potential by constructing a holomorphic function of chiral superfields, $W(\Phi)$, known as a superpotential. We then choose
\bea
\overbar{\La}_1^{\mathrm{SUSY}} &=& W(\Phi) \bigg|_{\theta \theta} + W(\Phi^\dagger)\bigg|_{\bar{\theta} \bar{\theta}} 
=F \frac{\p W}{\p A} + F^* \frac{\p W^*}{\p A^*} \, ,
\label{L1-susy}
\eea
where we have not yet specified the form of $W$. In order to do this, and complete the supersymmetrization of ${\bar{\cal{L}}}_{1}$, one must eliminate the auxiliary field $F$ using its equation of motion. We now address this issue, returning to the final component field expression for $\overbar{\La}_1^{\mathrm{SUSY}}$  at the end of the next subsection.

\subsection{Elimination of the $F$-field}

Let us first collect all those terms from the supersymmetric action that contain the complex auxiliary field $F$. Denoting this subset of the Lagrangian by $\overbar{\La}_F^{\mathrm{SUSY}}$, we find that
\bea
\overbar{\La}_F^{\mathrm{SUSY}} &=& 
F \frac{\p W}{\p A} + F^* \frac{\p W^*}{\p A^*} 
+ \bigg(\gamma + 2 \sqrt{2}\delta \hpi \bigg) F F^* 
\nn \\
&+& 
\bigg( - \frac{1}{2}c_3 -  c_4 -2 c_4 \hpi \bigg)
 \bigg(
- 2F F^* \Box \hpi
+ 2iF^* \p F \cdot \p \chi 
-  2i F \p F^* \cdot \p \chi
\bigg)
\nn \\
&+& 4 c_4 
\bigg( 
 (FF^*)^2 - \frac{1}{2} FF^* (\p \hpi)^2 - \frac{1}{2}FF^* (\p \chi)^2  
\bigg)
\nn \\
&+& \bigg( \frac{1}{8} c_2 + \frac{1}{3}c_3 - \frac{1}{3} c_4 + \big(\frac{2}{3}c_3 - \frac{4}{3}c_4 \big) \hpi \bigg)
\bigg( -   4 FF^* (\p \hpi)^2 - 4 FF^* (\p \chi)^2    +  4 (FF^*)^2 \bigg)
\nn \\
&-&
 \bigg( \frac{1}{4} c_4 + \frac{1}{6} c_4  \hpi \bigg)
\bigg(
- 4 \p_\mu (F \hpi_{,\nu}) \p^\mu (F^* \hpi^{,\nu})
- 4 \p_\mu (F \chi_{,\nu}) \p^\mu (F^* \chi^{,\nu})
+ 16 F F^* \p F \cdot \p F^*
\bigg)
\nn 
\eea
\bea
&+& 
\bigg( c_4 +  \frac{2}{3} c_4 \hpi \bigg)
\bigg(
- \hpi^{, \mu } \p_{\mu} (F \hpi_{, \nu}) F^{*, \nu} - \hpi^{, \mu } \p_{\mu} (F^* \hpi_{, \nu}) F^{, \nu}
- i \hpi^{, \mu } \p_{\mu} (F \chi_{, \nu}) F^{*, \nu} + i \hpi^{, \mu } \p_{\mu} (F^* \chi_{, \nu}) F^{, \nu}
\nn \\
&&
\qquad 
- \hpi^{, \mu } \p_\mu ( F \Box \hpi)F^* - \hpi^{, \mu} \p_{\mu} (F^* \Box \hpi)F
- \hpi^{, \mu } \p_\mu ( \p F \cdot \p \hpi) F^* - \hpi^{, \mu} \p_{\mu} (\p F^* \cdot \p \hpi)F
\nn \\
&&
\qquad 
- i \hpi^{, \mu } \p_\mu ( F \Box \chi)F^* + i \hpi^{, \mu} \p_{\mu} (F^* \Box \chi)F
+ i\hpi^{, \mu } \p_\mu ( \p F \cdot \p \chi) F^* - i \hpi^{, \mu} \p_{\mu} (\p F^* \cdot \p \chi)F
\nn \\
&&
\qquad 
- i \chi_{, \mu \nu} \p^{\mu}(F \hpi^{, \nu})F^* + i  \chi_{, \mu \nu} \p^{\mu}(F^* \hpi^{, \nu})F
+ \chi_{, \mu \nu} \p^{\mu}(F \chi^{, \nu})F^* +   \chi_{, \mu \nu} \p^{\mu}(F^* \chi^{, \nu})F
\nn \\
&&
\qquad 
- 4 F F^* \p F \cdot \p F^*
\bigg)
\nn \\
&+&
 \frac{32}{3} c_4 \bigg(
i \hpi^{, \mu} \p_{\mu} (F \hpi_{, \nu}) \chi^{, \nu} F^* - i \hpi^{, \mu} \p_{\mu} (F^* \hpi_{, \nu}) \chi^{, \nu} F
-  \hpi^{, \mu} \p_{\mu} (F \chi_{, \nu}) \chi^{, \nu} F^* - \hpi^{, \mu} \p_{\mu} (F^* \chi_{, \nu}) \chi^{, \nu} F
\bigg)
\nn \\
&-& 
\frac{32}{3} c_4 \bigg(
F F^* \hpi^{, \mu} \p_{\mu} (\p \hpi)^2 + F F^* \hpi^{, \mu} \p_\mu (\p \chi)^2 
\bigg)
+ \frac{64}{3} c_4 \bigg(
F (F^*)2 \p F \cdot \p \hpi + F^* F^2 \p F^* \cdot \p \hpi
\bigg)
\nn \\
&+&
 \bigg( - \frac{19}{6} c_4 - \frac{76}{9}c_4 \hpi \bigg) 
 \bigg(
 - 2 F F^* ( (\p \hpi)^2 + (\p \chi)^2 ) \Box \hpi
 + 4 (F F^* )^2 \Box \hpi
 \bigg)
\nn \\
&+&
\bigg( -c_3 - \frac{11}{3} c_4 + (- \frac{2}{3} c_3 - \frac{88}{9}c_4) \hpi  \bigg)
\bigg(
- \p_\mu (F F^* (\p \hpi)^2 ) \hpi^{, \mu} 
- \p_\mu (F F^* (\p \chi)^2 ) \hpi^{, \mu} 
+ \p_\mu(F F^*) \hpi^{, \mu}
\bigg)
\nn \\
&+&
\bigg(
-\frac{1}{16}c_2 - \frac{1}{3} c_3 - \frac{9}{4} c_4 
+ (-\frac{1}{24} c_2 -\frac{8}{9}c_3 - \frac{21}{2} c_4) \hpi
\bigg)
\nn \\
&&\qquad \qquad \qquad \qquad 
\bigg( 
(- 4 F F^* \big( (\p \hpi)^2 - (\p \chi)^2\big)
+ 4 (F F^*)^2 ) \big( (\p \hpi)^2 + (\p \chi)^2\big)
\bigg) \, ,
\nn \\
\label{LF-susy}
\eea
where, for convenience,  we have defined two parameters, $\gamma$ and $ \delta$, of mass dimension 2 as
\bea
\gamma \equiv  \frac{(c_2 + 2 c_3 + \frac{4}{3} c_4)}{\alpha^2} \, ,
\qquad 
\delta \equiv \frac{1}{\sqrt{2}} \frac{(- \frac{1}{3}c_2 + \frac{4}{3}c_3 + \frac{20}{9} c_4)}{\alpha^2} \, .
\label{gamma-delta}
\eea
As mentioned previously, this Lagrangian not only contains terms that are cubic or higher order polynomials in $F$ and $F^*$, but also terms which involve derivatives of $F$; including  terms with two derivatives on $F$, such as $F F^* \p F \cdot \p F^*$. The question then arises as to whether or not $F$ can legitimately be considered an auxiliary field--which can be eliminated via an algebraic equation of motion--or is, instead, a dynamical scalar which propagates in the same manner as $\hpi$ and $\chi$. This issue is typical of higher derivative theories of supersymmetry and supergravity--see, for example, \cite{Khoury:2011da, Koehn:2012ar}. Of course, a propagating complex scalar $F$ is not necessarily a problem for supersymmetry, since it can be associated with two extra propagating degrees of freedom in the Weyl spinor $\psi$.

In \cite{Deen:2017jqv}, we presented a method for addressing this issue in the case of the supersymmetric conformal Galileons. We now adapt this method to the supersymmetric heterotic Lagrangians. To begin, we observe that after restoring $\alpha$ in the supersymmetric Lagrangians given above, the mass dimension 1 scalar $F$ always appears in the ratio $F / \alpha$. This mirrors the structure of the derivative $\p \hpi$, which always appears in the form $\p \hpi / \alpha$. Since we are restricting the derivative terms to be small so as to limit the derivative expansion to the four heterotic Galileons discussed above, it is natural to demand that in the supersymmetric extension, $F/ \alpha$, be small as well. To be explicit, we henceforth require that
\bea
\bigg|\frac{F}{\alpha} \bigg|^2 \ll 1 \, .
\label{HD2}
\eea
This condition means that \eqref{LF-susy} is composed of terms which are suppressed by successively higher powers of $\alpha^2$, as were the heterotic Galileons in \eqref{heterotic-galileons}. Therefore, higher order terms  in $F$ and those involving derivatives of $F$, which arise in the supersymmetrization of $\bar{\La}_3$ and $\bar{\La}_4$, will be small compared to the linear and quadratic terms from $\overbar{\La}_1^{\mathrm{SUSY}}$ and $\overbar{\La}_2^{\mathrm{SUSY}}$. This allows us to treat $F$ as an auxiliary field, since the terms that would ``propagate" it are heavily suppressed. We can then solve for $F$ perturbatively and substitute the result into Lagrangian \eqref{LF-susy}. The perturbative expansion for $F$ will be of the form
\be
F=F^{(0)}+F^{(1)}+\dots \ ,
\label{cup1}
\ee
where $F^{(0)}$ arises from solving the $F$ equation of motion using $\overbar{\La}_1^{\mathrm{SUSY}}$ and $\overbar{\La}_2^{\mathrm{SUSY}}$ only, $F^{(1)}$ is then computed by adding the contribution of $\overbar{\La}_3^{\mathrm{SUSY}}$ to the $F$ equation of motion, and so on. Let us write \eqref{cup1} in the form.
\be
F=F^{(0)}(1+\frac{F^{(1)}}{F^{(0)}}+\dots) \ .
\label{cup12}
\ee
It is clear from the above discussion that 
\be
\frac{F^{(1)}}{F^{(0)}} \sim (\frac{\partial}{\alpha})^{2} \ll 1 
\label{cup3}
\ee
and, hence, $F$ is very well approximated by $F^{(0)}$. Therefore, in the remainder of this paper we will always  take $F=F^{(0)}$ and ignore higher order corrections.
In doing so, it will become clear that the coefficients $c_i$, which arose from the construction of the DBI action \eqref{DBI-heterotic}, can no longer be arbitrary and must satisfy certain specific constraints. As stated above, the largest terms in \eqref{LF-susy} arise from  $\overbar{\La}_1^{\mathrm{SUSY}}$ and $\overbar{\La}_2^{\mathrm{SUSY}}$, and are given by
\bea
\overbar{\La}_{F}^{\mathrm{SUSY}(0)} = F^{(0)}  \frac{\p W}{\p A} + F^{*(0)}  \frac{\p W^*}{\p A^*} + \big( \gamma + 2 \sqrt{2} \delta \hpi \big) F^{(0)}  F^{*(0)} \, ,
\label{LF-0}
\eea
where the dimension 2 constants $\gamma$ and $\delta$ are the linear combinations of the coefficients $c_i$ given in \eqref{gamma-delta}. Solving the equation of motion for $F^{(0)}$, we find that
\bea
F^{(0)} = - \frac{1}{\big( \gamma + 2 \sqrt{2} \delta \hpi \big)} \frac{\p W^*}{\p A^*} \, .
\label{F0-1}
\eea
For the holomorphic function, $W(A)$, we choose
\bea
W(A) = \beta_1 A + \beta_2 A^2  \, ,
\label{W-1}
\eea
where the constant coefficients $\beta_1$, $\beta_{2}$ each have mass dimension 3. Furthermore, it will be sufficient to take each of the $\beta_{i}$ coefficients to be real numbers. As we will demonstrate below, superpotential \eqref{W-1} leads to the correct scalar Lagrangian ${\bar{\cal{L}}}_{1}$ presented in \eqref{heterotic-galileons}--and appears to be the minimal holomorphic superpotential which can do so. Hence, although more complicated superpotentials might be possible, we will, for simplicity, take $W(A)$ to be the quadratic function given in \eqref{W-1}. It then follows from \eqref{F0-1} and \eqref{W-1} that, to linear order in $\hpi$,  
\bea
F^{(0)} &=&
 -\frac{\beta_1}{\gamma} + \sqrt{2} \bigg( 2 \frac{\beta_1}{\gamma} \bigg(\frac{\delta}{\gamma}\bigg) - \frac{\beta_2}{ \gamma} \bigg) \hpi
+ i \sqrt{2} \frac{\beta_2}{\gamma} \chi
- i4\frac{\beta_2}{\gamma} \bigg(\frac{\delta}{\gamma}\bigg)  \hpi \chi \, .
\nn \\
\label{F0-2}
\eea
Note from the denominator in \eqref{F0-1} that to consistently work to first order in $\hpi$ only, we have to constrain $\delta$ and $\gamma$ to satisfy
\be
|\frac{\delta}{\gamma}| \lesssim 1 \ .
\label{H1}
\ee
It then follows from \eqref{gamma-delta} that the coefficients $c_{i}$, $i=1,2,3,4$ must satisfy the constraint that
\be
|-\frac{1}{3}c_{2}+\frac{4}{3}c_{3}+\frac{20}{9}c_{4}| \lesssim \sqrt{2}|c_{2}+2c_{3}+\frac{4}{3}c_{4}| \ .
\label{homea}
\ee

Before discussing the conditions under which $|F^{(0)}/\alpha|^{2} \ll 1$, one must first compute $\overbar{\La}_{F}^{\mathrm{SUSY}(0)}$ and determine whether or not it is consistent with 
${\bar{\cal{L}}}_{1}$ in \eqref{heterotic-galileons}. Putting expression \eqref{F0-2} into \eqref{LF-0}, we find that
the complete scalar potential energy is given by
\bea
V(\hpi, \chi) &=& - \overbar{\La}_{F}^{\mathrm{SUSY}(0)} 
\nn\\
&=& 
\frac{\beta_1^2}{\gamma} + \frac{2\sqrt{2}}{\gamma^2} \big( - \beta_1^2 \delta +   \beta_1 \beta_2 \gamma \big) \hpi
+ \frac{2}{\gamma} \beta_2^2  \chi^2 
\nn \\
&-&
\frac{4\sqrt{2}}{\gamma^2} \beta_2^2 \delta \hpi \chi^2 \, .
\label{V}
\eea
Setting $\chi=0$ in expression $\eqref{V}$, and demanding that the result reproduce $-\bar{\La}_1$ in \eqref{heterotic-galileons} exactly, necessitates the imposition of two constraints on  $\beta_1$ and $\beta_2$. These are
\bea
\beta_1^2 = \gamma \bigg(\frac{3}{10}c_1 + c_2 + \frac{4}{3}c_3 \bigg)\, ,
\qquad
- \beta_1^2 \delta + \beta_1 \beta_2 \gamma = - \frac{\gamma^2}{2\sqrt{2}}\bigg(c_1 + \frac{4}{3}c_2 - \frac{8}{9} c_3\bigg) \, .
\label{beta-eqn}
\eea
Note that the first constraint immediately implies that 
\be
 \gamma \bigg(\frac{3}{10}c_1 + c_2 + \frac{4}{3}c_3 \bigg) > 0 \ .
\label{homeb}
\ee
We conclude that choosing the quadratic superpotential \eqref{F0-1} leads to the appropriate $N=1$ supersymmetrization $\overbar{\La}_{F}^{\mathrm{SUSY}(0)}$ of the scalar 
${\bar{\cal{L}}}_{1}$ heterotic Galileon as long as the two coefficients $\beta_{i}$, $i=1,2$ of $W(A)$ satisfy the constraints in \eqref{beta-eqn}.

Having determined this, we must now ensure that $F^{(0)}$ presented in \eqref{F0-2} satisfies the constraint given in \eqref{HD2}; that is, that $|F^{(0)} / \alpha|^{2} \ll 1$. It is clear from expression \eqref{F0-2} and \eqref{H1} that this will be the case as long as
\be
|\frac{{\beta}_{i}}{\gamma}| \ll 1, \quad i=1,2  \ .
\label{HD2a}
\ee
Solving these inequalities subject to the constraints given in \eqref{beta-eqn}, leads to two conditions on the coefficients $c_{i}$, $i=1,2,3,4$. First, demanding that $|\beta_{1}/\gamma| \ll 1$ and using the first expression in \eqref{beta-eqn}, leads to the inequality
\be
\frac{3}{10}c_{1}+c_{2}+\frac{4}{3}c_{3} \ll c_{2}+2c_{3}+\frac{4}{3}c_{4} \ .
\label{HD3}
\ee
Second, the constraint $|\beta_{2}/\gamma| \ll 1$ and the second expression in \eqref{beta-eqn} implies that
\be
|-c_{1}-\frac{4}{3}c_{2}+\frac{8}{9}c_{3}| \ll 2\sqrt{2}| \beta_{1}| \ .
\label{HD4}
\ee

Before continuing, we note that having chosen the form of the superpotential $W(A)$ in \eqref{W-1}, one can now write the expression for $\overbar{\La}_1^{\mathrm{SUSY}}$ in \eqref{L1-susy} in terms of component fields. It is given by
\be
\overbar{\La}_1^{\mathrm{SUSY}}=\beta_{1}(F+F^{*})+ \sqrt{2}\beta_{2}(F+F^{*})\hpi +i\sqrt{2}\beta_{2}(F-F^{*})\chi \ .
\label{car2}
\ee

\subsection{Physical Requirements}
As discussed in a previous paper \cite{Deen:2017jqv}, we would like our Lagrangian to be such that it admits a solution of the equations of motion for which, if we take both $\chi=0$ and $\p_\mu \chi=0$ initially, then $\chi= \p_\mu \chi=0$ remains unchanged as time evolves. That is, any dynamical motion is purely in the $\hpi$ direction in field space.  This requires an analysis of the potential \eqref{V}. For this to be the case, it is necessary that
\bea
m_{\chi}^2 = \frac{\p^2 V}{\p \chi^2} \bigg|_{\chi = 0} \geq 0 
\label{chi-mass} 
\eea
for all values of $\hpi$, where
\bea
\frac{\p^2 V}{\p \chi^2} \bigg|_{\chi = 0} &=&
\frac{4\beta_{2}^{2}}{\gamma}\big(1-2\sqrt{2}(\frac{\delta}{\gamma})\hpi \big) \, .
\eea
Using \eqref{H1}, it follows that \eqref{chi-mass} will be satisfied as long as one chooses the $c_{i}$ coefficients such that $\gamma > 0$. It then follows from \eqref{gamma-delta} that
\be
c_2 + 2c_3 + \frac{4}{3} c_4 > 0
\label{homec}
\ee
Note that putting this result back in \eqref{homeb}, simplifies that constraint to become
\be
\frac{3}{10}c_1 + c_2 + \frac{4}{3}c_3 > 0
\label{homed}
\ee
Of course, condition \eqref{chi-mass} will lead to the solution $\chi=\partial_{\mu}\chi=0$ only if one assumes a non-ghost like kinetic energy for $\chi$. To ensure that this is the case, let us combine the kinetic terms for $\chi$ which arise from \eqref{L2-susy} and \eqref{LF-susy}. That is
\bea
&&
\bigg[- \frac{1}{2} c_2 -  c_3  - \frac{2}{3}c_4+ ( \frac{1}{3}c_2 - \frac{4}{3}c_3 - \frac{20}{9} c_4) \hpi 
\nn \\
&&+ \bigg( - \frac{1}{2} c_2 - \frac{4}{3} c_3 - \frac{2}{3} c_4 + (-\frac{8}{3} c_3 + \frac{16}{3} c_4) \hpi \bigg)
F^{(0)}F^{*(0)}
\nn \\
&&+ \bigg( - \frac{1}{4} c_2 - \frac{4}{3} c_3 - 9 c_4 + (- \frac{1}{6}c_2 -\frac{32}{9} c_3 - 42 c_4) \hpi \bigg)
(F^{(0)}F^{*(0)})^2 \bigg] (\p \chi)^2  \,,
\label{chi-KE}
\eea
where $F^{(0)}$ is given by \eqref{F0-2}. 
This kinetic energy will be ghost free if and only if the coefficient of $(\partial\chi)^{2}$ is negative for any value of $\hpi$. It follows that the $c_{i}$ coefficients must satisfy
\be
\big(- \frac{1}{2} c_2 -  c_3  - \frac{2}{3}c_4\big)+\big( - \frac{1}{2} c_2 - \frac{4}{3} c_3 - \frac{2}{3} c_4  \big)|F^{(0)}|^{2}+ \big( - \frac{1}{4} c_2 - \frac{4}{3} c_3 - 9 c_4  \big)|F^{(0)}|^{4} < 0
\label{hang1}
\ee
and
\be
\big(\frac{1}{3}c_2 - \frac{4}{3}c_3 - \frac{20}{9} c_4\big)+\big(-\frac{8}{3} c_3 + \frac{16}{3} c_4 \big)|F^{(0)}|^{2}+ \big( - \frac{1}{6}c_2 -\frac{32}{9} c_3 - 42 c_4  \big)|F^{(0)}|^{4} < 0
\label{hang2}
\ee
This imposes two additional extra conditions on the coefficients $c_i$. 
It is important to note that the kinetic energy term for $\hpi$ is identical to that of $\chi$; one simply replaces $(\p \chi)^2$ in equation \eqref{chi-KE} with $(\p \hpi)^2$. By requiring $\chi$ be ghost free, we thus ensure that $\hpi$ is ghost-free as well; that is, requiring $(\partial\hpi)^{2}$ to be ghost free imposes no additional constraints.

Finally, there are two additional ``physical'' constraints that we impose on the supersymmetric three-brane action. The first is that we require the three-brane ``tension'' to be positive. It is straightforward to show that this will be the case if and only if 
\be
c_{2} > 0 \ .
\label{hang3}
\ee
Second, on physical grounds we would like the three-brane to be attracted to the {\t observable} orbifold plane by the potential energy in the worldvolume action. It then follows from
\eqref{V} that
\be
-c_{1}-\frac{4}{3}c_{2}+\frac{8}{9}c_{3} > 0
\label{hang4}
\end{equation}
Note that this simplifies \eqref{HD4} to become
\be
-c_{1}-\frac{4}{3}c_{2}+\frac{8}{9}c_{3} \ll 2\sqrt{2}| \beta_{1}| \ .
\label{hang5}
\ee

\subsection{Summary of Constraints on the Coefficients}

The complete set of constraints on the coefficients $c_{i}$, $i=1,2,3,4$ determined above are the following:

\begin{enumerate}
\item  $|-\frac{1}{3}c_{2}+\frac{4}{3}c_{3}+\frac{20}{9}c_{4}| \lesssim \sqrt{2}(c_{2}+2c_{3}+\frac{4}{3}c_{4})$
\item  $ \frac{3}{10}c_1 + c_2 + \frac{4}{3}c_3  > 0$
\item  $\frac{3}{10}c_{1}+c_{2}+\frac{4}{3}c_{3} \ll c_{2}+2c_{3}+\frac{4}{3}c_{4}$
\item  $-c_{1}-\frac{4}{3}c_{2}+\frac{8}{9}c_{3} \ll 2\sqrt{2}| \beta_{1}|$
\item  $c_2 + 2c_3 + \frac{4}{3} c_4 > 0$
\item  $c_{2} > 0$
\item  $-c_{1}-\frac{4}{3}c_{2}+\frac{8}{9}c_{3} > 0$
\item  $\big(- \frac{1}{2} c_2 -  c_3  - \frac{2}{3}c_4\big)+\big( - \frac{1}{2} c_2 - \frac{4}{3} c_3 - \frac{2}{3} c_4  \big)|F^{(0)}|^{2}+ \big( - \frac{1}{4} c_2 - \frac{4}{3} c_3 - 9 c_4  \big)|F^{(0)}|^{4} < 0$
\item  $\big(\frac{1}{3}c_2 - \frac{4}{3}c_3 - \frac{20}{9} c_4\big)+\big(-\frac{8}{3} c_3 + \frac{16}{3} c_4 \big)|F^{(0)}|^{2}+ \big( - \frac{1}{6}c_2 -\frac{32}{9} c_3 - 42 c_4  \big)|F^{(0)}|^{4} < 0$ 
\end{enumerate}
where $\beta_{1}$ is defined in \eqref{beta-eqn}.

In this paper, we will make no attempt to present a complete set of solutions to these conditions.  
Instead, we will perform a numerical scan over four dimensional $c_i$-space to demonstrate that there exist reasonable values of the coefficients $(c_1, c_2, c_3, c_4)$ which satisfy the constraints given above. We will restrict the values of the $c_i$'s so that, in units of $\alpha =1$, their absolute value is bounded by 
\bea
 | c_i | \leq 10 \, ,
\eea
which is a physically realistic assumption. We first set up a four-dimensional grid with an incremental step size $\Delta c $, and evaluate every point in the grid to see if they satisfy the required inequalities. Let us be precise about the numerical definition of the $\ll$ symbol appearing in inequalities 3. and 4. in the summary of constraints. We will, in this analysis, take it to mean that the ratio of the two quantities given is less than $1/25$; that is
\bea
a \ll b \Rightarrow \frac{a}{b} < \frac{1}{25} \ .
\eea
Furthermore, we will work in a restricted region of \emph{field} space, such that the magnitude of both $\hpi$ and $\chi$ cannot exceed $1/10$. Note that for $\hpi$ this is consistent with condition \eqref{VT2}. This enables us to evaluate inequalities involving the field dependent quantity $F^{(0)}$. We do so by replacing $\hpi$ and $\chi$ with their maximum value, that is,  $1/10$, in the expression for $F^{(0)}$.

A preliminary search reveals that $c_1$ must be negative for there to be any satisfactory points. Taking $c_1$ to be a {\it fixed} negative value,  we perform a more refined scan over the remaining $c_i$'s by taking $\Delta c = 0.01$. The results for $c_{1}=-1$ and $c_{1}=-10$ are presented in Figures 1 and 2 respectively. We note that taking $c_1$ to be more negative, as in Figure 2, means that more points can satisfy the constraints. This is clear from the larger ``volume" of valid points in Figure 2, as opposed to those of Figure 1. 
Finally, we find that $|F^{(0)}|^2$ is indeed small for all of the valid points displayed in Figures 1 and 2; as is required for the perturbative expansion of $F$ to be valid.  At its largest, $|F^{(0)}|^2 \simeq 0.04$ in both cases, but is generically smaller, as can be seen in Figure 3.

\begin{figure}[h!]
\begin{center}
\includegraphics[scale=0.8]{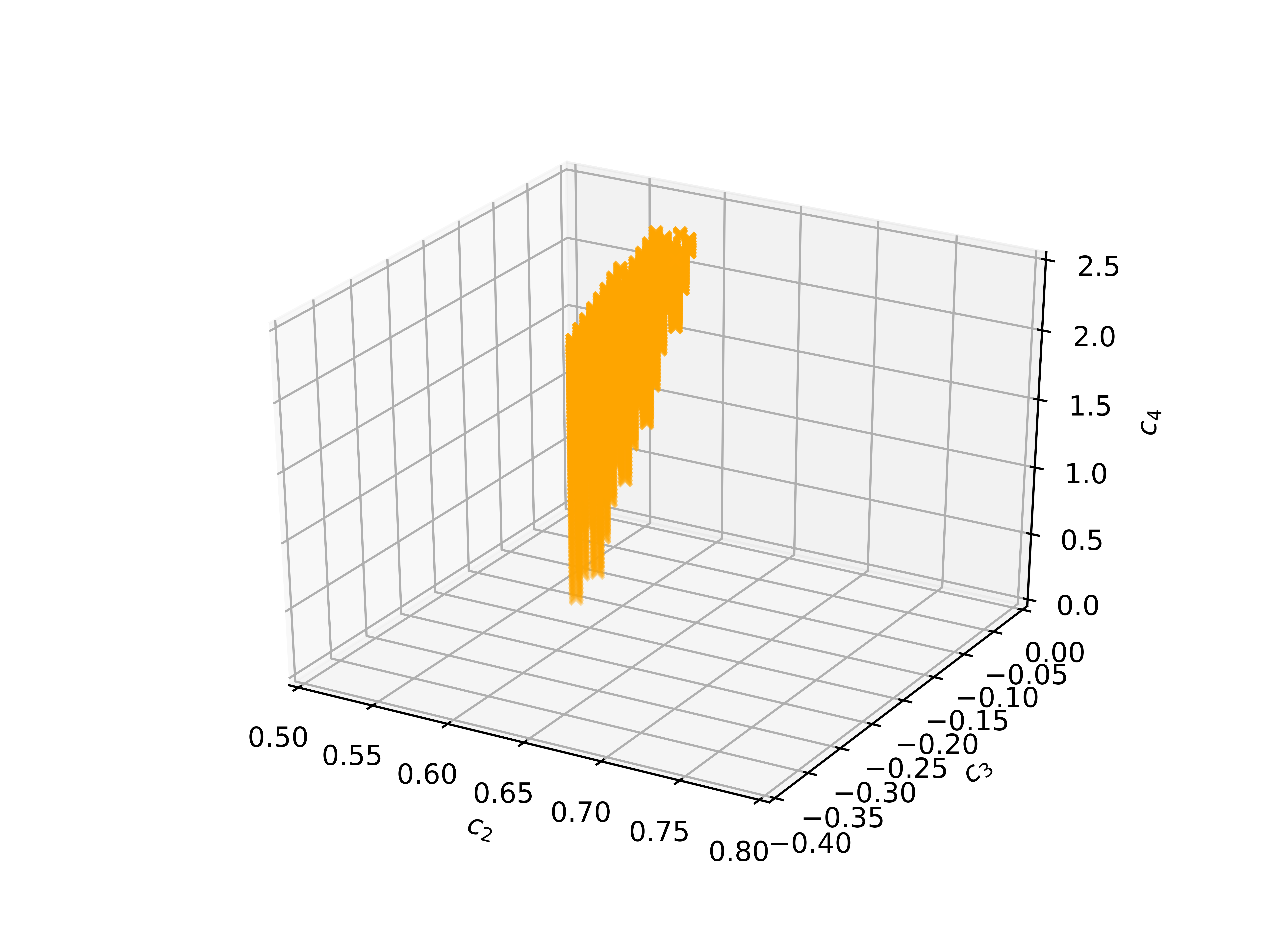}
\caption{\small{Numerical scan over $-10 \leq c_{i} \leq 10$ for $i=2,3,4$, taking $c_1 = -1.0$ and with step size $\Delta c = 0.01$. Points which satisfy all inequalities in the ``summary of constraints'' are labelled by an orange $\times$.} }
\label{c1-one}
\end{center}
\end{figure}
\begin{figure}[!h]
\begin{center}
\includegraphics[scale=0.8]{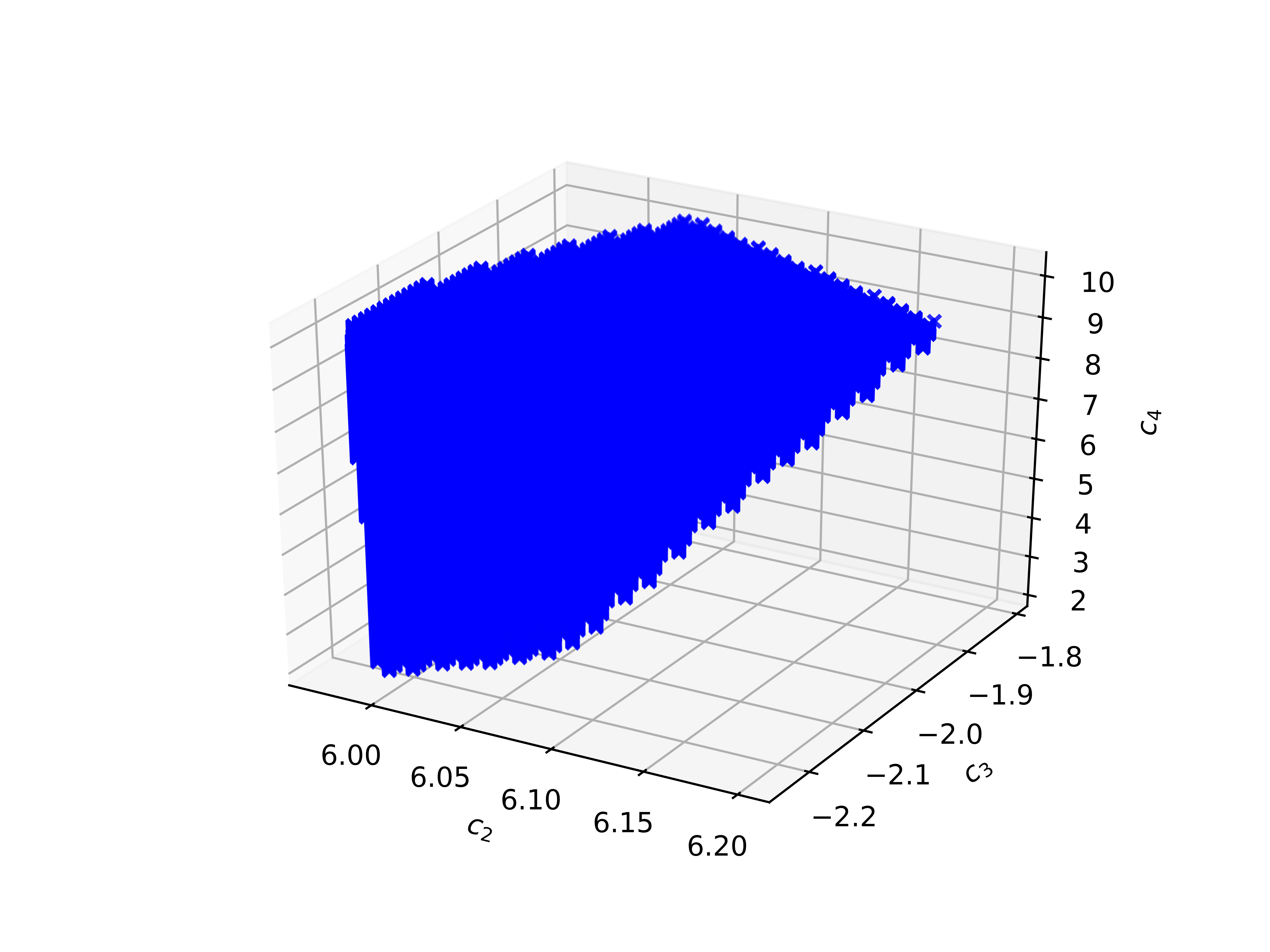}
\caption{\small{Numerical scan over $-10 \leq c_{i} \leq 10$ for $i=2,3,4$, taking $c_1 = -10.0$ and with step size $\Delta c = 0.01$. Points which satisfy all inequalities given in the ``summary of constraints'' are labelled by a blue $\times$.}}
\label{c1-ten}
\end{center}
\end{figure}
\begin{figure}[h]
\begin{subfigure}[h]{0.5\linewidth}
\includegraphics[width=\linewidth]{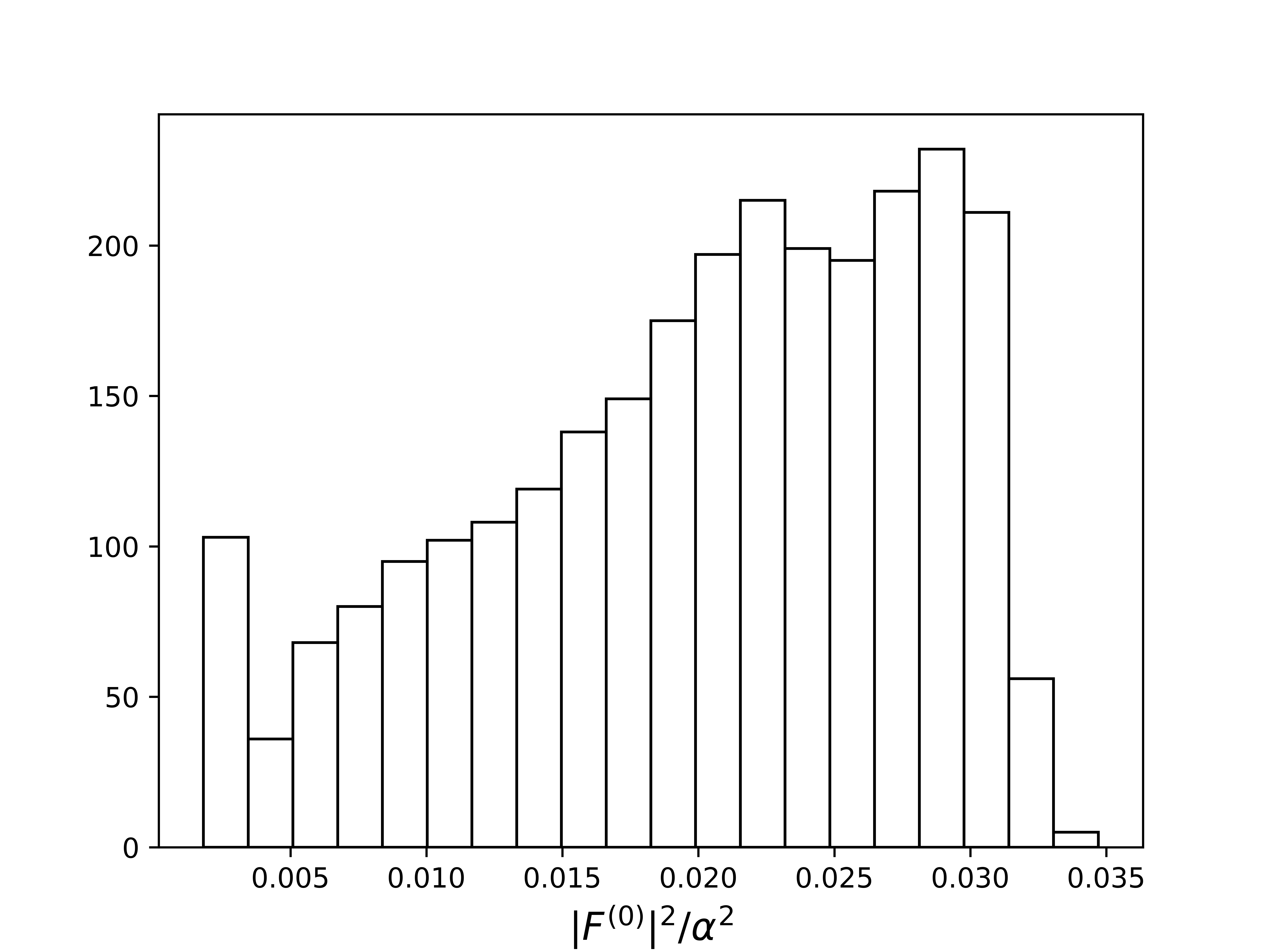}
\caption{}
\end{subfigure}
\hfill
\begin{subfigure}[h]{0.5\linewidth}
\includegraphics[width=\linewidth]{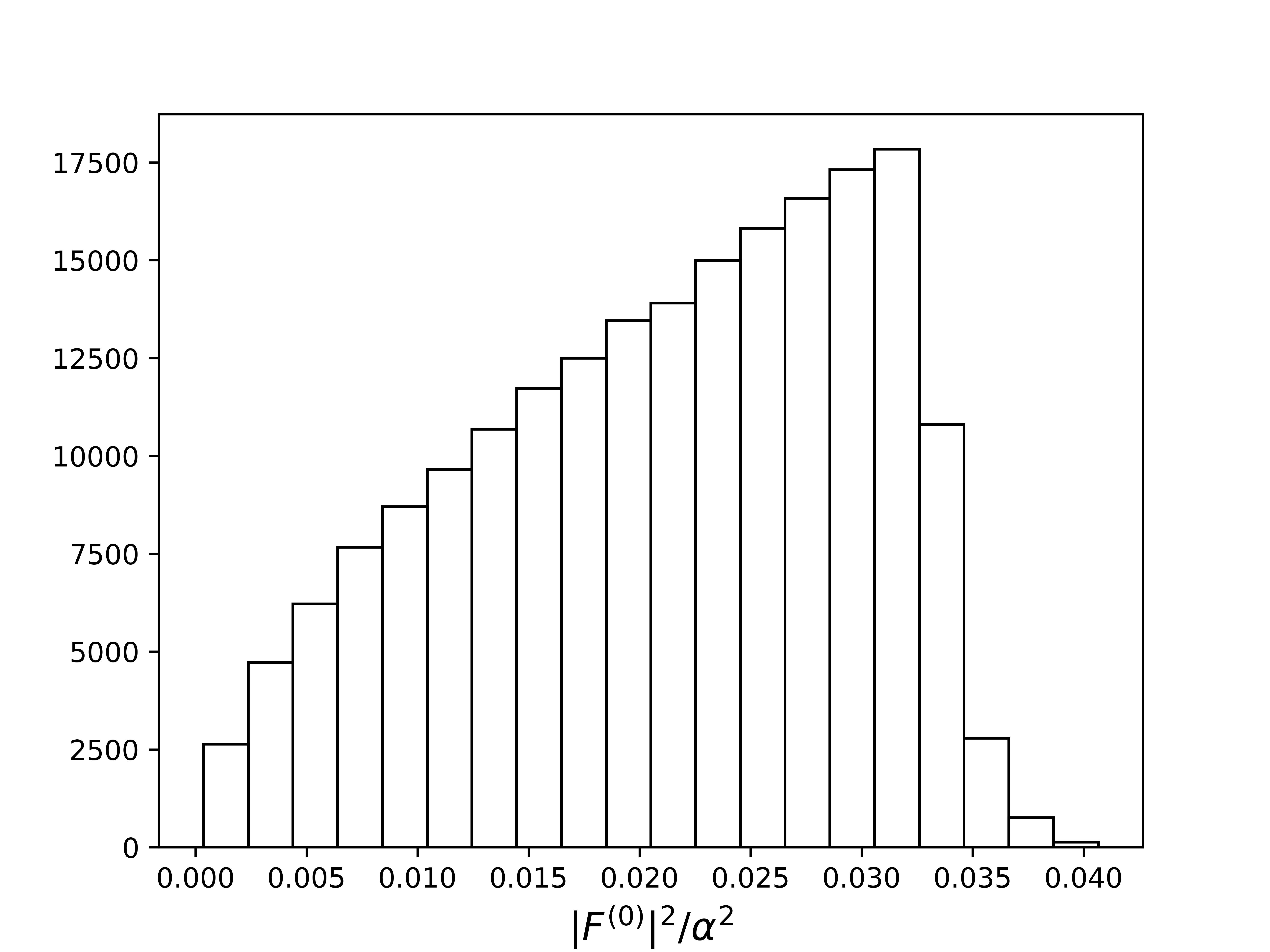}
\caption{}
\end{subfigure}%
\label{F-histogram}
\caption{\small{Histograms of $|F^{(0)}|^2 / \alpha^2$ for (a)  $c_1= -1$ and (b) $c_1 = -10$. The data are from the same numerical scans as in Figures 1 and 2. Plots (a) and (b) each represent a total of 2,701 and 198,903 points respectively.}}
\end{figure}
For completeness, the entire Lagrangian density for the $N=1$ flat superspace heterotic Galileons, ${\overbar{\La}}^{\mathrm{SUSY}}_{1+2+3+4}$, is presented in component fields in Appendix A. All terms containing the chiral fermion $\psi$ have been set to zero. The Lagrangian, after the elimination of $F$ and subject to the various constraints discussed above, can be obtained by substituting $F \rightarrow F^{(0)}$ and using only those coefficients $c_{i}$,  $i=1,2,3,4$ satisfying the conditions presented in Subsection 4.7.


\section{Extension of Heterotic Galileons to $N=1$ Supergravity}
\label{sec-5}
We now proceed to extend the heterotic Galileons in equation \eqref{heterotic-galileons} to local supersymmetry; that is, to $N=1$  supergravity. This is essential if we are to explore the cosmological implications of three-branes in heterotic $M$-theory. Much of the computation involved is similiar to that presented in \cite{Deen:2017jqv},  so we will limit ourselves to the most pertinent details. We continue to use the superspace formalism described in \cite{Wess}, where the global anti-commuting $\theta^\alpha$ coordinates are now replaced by local superspace coordinates $\Theta^\alpha$. These now define the superfield expansions. As above, we will embed the real scalar field $\hpi$ appearing in \eqref{heterotic-galileons} in a complex scalar field $A = \frac{1}{2}(\hpi + \chi)$, which is taken to be the lowest component of a chiral superfield
\bea
\Phi = A + \sqrt{2} \Theta \psi + \Theta \Theta F \ .
\eea
As in the flat superspace case, this chiral superfield contains, in addition to $A$, a two component Weyl spinor $\psi$ and a complex scalar auxiliary field $F$.
With the exception of the supergravity extension of ${\bar{{\cal{L}}}}_{1}$, our $N=1$ locally supersymmetric Lagrangians are all of the form
\bea
\int d^2 \Theta \, 2\mathcal{E} (\oD^2 - 8R) \mathcal{O}(\Phi, \Phi^\dagger) + \mathrm{h.c.} \, ,
\eea
where $\mathcal{O}(\Phi, \Phi^\dagger) $ is a Lorentz scalar involving $\Phi$ and $\Phi^\dagger$.
The integral is over half of superspace, where the chiral projection operator $\oD^2 - 8R$, involving the curvature {\it chiral superfield} $R$, acts on $\mathcal{O}$ so as to make the combination $(\oD^2 - 8R) \mathcal{O}$ a chiral superfield. The geometrical {\it chiral density} $\mathcal{E}$ ensures that the Lagrangian has the appropriate transformation properties under local $N=1$ supersymmetry.

Local $N=1$ supersymmetry necessitates the introduction of a supergravity multiplet containing the spin 2 graviton $e_\mu^{~a}$  and the spin 3/2 gravitino $\psi_\mu^{~\alpha}$. However, the off-shell superspace formalism that we are using requires the addition of two new auxiliary fields, a complex scalar $M$ and a real vector field $b_\mu$, to the supergravity multiplet. Both $M$ and $b_\mu$ have mass dimension one, and appear in the $\Theta$ expansions of the chiral superfield $R$ and the geometrical chiral density $\mathcal{E}$. Explicitly, one finds that
\bea
R &=&
-\tfrac{1}{6}M 
+\Theta^2 
(\tfrac{1}{12}\mathcal{R} - \tfrac{1}{9}M M^* - \tfrac{1}{18}b^\mu b_\mu + \tfrac{1}{6}i e_a^{~\mu} \D_\mu b^a)
\nn \\
\mathcal{E} &=& \tfrac{1}{2}e    -\tfrac{1}{2} \Theta^2e M^* \ ,
\label{bw3}
\eea
where $\mathcal{R}$ is the four-dimensional Ricci scalar (not to be confused with the similar notation for the radius of curvature in the $AdS_{5}$ case) and $e=\mathrm{det} \, e^{~a}_{\mu}$.
For more details on the construction of $N=1$ supergravity Lagrangians using the superspace formalism, we refer the reader to \cite{Deen:2017jqv, Wess}, as well as to \cite{ Koehn:2012ar,Ciupke:2015msa,Baumann:2011nm}. Higher-derivative Lagrangians in $N=1$ supergravity have been examined in \cite{ Koehn:2012ar,Ciupke:2015msa,Baumann:2011nm, Farakos:2012je, Farakos:2012qu,Farakos:2013zya,Ciupke:2016agp}.
Therefore, in addition to the auxiliary field $F$ of the chiral supermultiplet $\Phi$, one must now examine the behaviour of the supergravity auxiliary fields $M$ and $b_{\mu}$. 

Let us first consider the supergravity extension of $\bar{\La}_1$, $\bar{\La}_2$ and $\bar{\La}_3$, deferring the discussion of $\bar{\La}_4$ to the end of this section. For simplicity, we will set $\alpha =1$ unless otherwise stated. However, to explicitly demonstrate where effects due to gravitation arise, we will exhibit the factors of the Planck mass, $M_P$, wherever they occur in our expressions. As above, we will not present terms which involve fermions, since these are not relevant for this paper. Therefore, in addition to dropping terms involving the Weyl fermion $\psi$, we will also exclude terms which containing the gravitino $\psi_\mu^{~\alpha}$. The appropriate supergravity extensions of  $\bar{\La}_1$ and $\bar{\La}_2$ are given, in terms of superfields, by

\bea
\overbar{\La}_1^{\mathrm{SUGRA}} &=& \int d^2 \Theta \, 2 \mathcal{E} \, W(\Phi) + \mathrm{h.c.} \label{air1}
\\
\nn \\
\bar{\La}_2^{\mathrm{SUGRA}} 
&=& M_P^2 \int d^2 \Theta \, 2 \mathcal{E} \bigg( - \frac{3}{8} (\oD^2 - 8R ) e^{-K(\Phi , \Phi^\dagger) /3 M_P^2}\bigg) + \mathrm{h.c} \label{air2}
\eea
To be consistent with the flat supersymmetry results of the previous section, it follows from \eqref{p1} and \eqref{W-1} that one must take
\bea
K(\Phi , \Phi^\dagger) = \gamma \Phi \Phi^\dagger + \delta (\Phi^2 \Phi^\dagger + \Phi (\Phi^\dagger)^2) \, ,
\qquad
W(\Phi) = \beta_1 \Phi + \beta_2 \Phi^2 \, ,
\eea
where $\gamma, \delta$ are defined in equation \eqref{gamma-delta} and $\beta_1$, $\beta_{2}$ are real coefficients. Written in terms of
components fields, we find that \eqref{air1} and \eqref{air2} become
\bea
\frac{1}{e}\bar{\La}_1^{\mathrm{SUGRA}} &=&
\frac{\p W}{\p A} F + \frac{\p W^*}{\p A^*} F^* - W M^* - W^* M \ ,
\nn \\
\nn \\
\frac{1}{e} \bar{\La}_2^{\mathrm{SUGRA}} 
&=& 
M_P^2  e^{-\frac{1}{3} \Km } \bigg(- \frac{1}{2} \mathcal{R} - \frac{1}{3} M M^* + \frac{1}{3} b^\mu b_\mu \bigg)
+ 3 M_P^2 \frac{\p^2 e^{-\frac{1}{3} \Km }  }{\p A \p A^*} (\p A \cdot \p A^* - F F^*)
\nn \\
&+& i M_P^2 b^\mu (\p_\mu A \frac{\p e^{-\frac{1}{3} \Km } }{\p A} - \p_\mu A^* \frac{\p e^{-\frac{1}{3} \Km } }{\p A^*})
+ M_P^2 (MF\frac{\p e^{-\frac{1}{3} \Km } }{\p A} + M^* F^* \frac{\p e^{-\frac{1}{3} \Km } }{\p A^*})
\nn \\
\eea
respectively.

The extension to $N=1$ supergravity of $\bar{\La}_3$ is constructed from two terms, 
\bea
\overbar{\La}_{3,\mathrm{I}} &=& 
- \frac{1}{64} \int d^2 \Theta \, 2 \m{E}\,(\oD^2 - 8 R )\bigg(- \frac{1}{\sqrt{2}} c_3 - \frac{ \sqrt{2}}{3}c_4  - 2 c_4(\Phi + \Phi^\dagger) \bigg) \D \Phi \D \Phi \oD^2 \Phi^\dagger + \mathrm{h.c.}
\nn \\
\nn 
&=& 
\bigg( -\frac{1}{\sqrt{2}} c_3 - \frac{\sqrt{2}}{3}c_4 - 2c_4(A + A^*)\bigg)
\bigg(
(\p A)^2 \big( \nabla_\mu \partial^\mu A^* + \frac{2}{3} i b^\mu \p_\mu A + \frac{2}{3} M^* F^* \big)
\nn \\ 
&& \qquad \qquad 
+
(\p A^*)^2 \big( \nabla_\mu \partial^\mu A + \frac{2}{3} i b^\mu \p_\mu A^* + \frac{2}{3} M F \big)
- \frac{4}{3} M F^2 F^* - \frac{4}{3} M^* (F^*)^2 F 
\nn \\
 \qquad 
&& \qquad \qquad \qquad 
+ 2 F^* \p F \cdot \p A + 2 F \p F^* \cdot \p A^* 
+ \frac{1}{6} i F F^* b^\mu( \p_\mu A - \p_\mu A^*)
\bigg)   
\nn 
\eea
\bea
&+&2c_4
\bigg(  2(F F^*)^2 - 4 F F^* \p A \cdot \p A^* -  F F^*\big( (\p A)^2 + (\p A^*)^2 \big)  \bigg)
\nn \\
&-&
\bigg( -\frac{1}{\sqrt{2}} c_3 - \frac{\sqrt{2}}{3}c_4 - 2c_4(A + A^*) \bigg) 
\bigg( (\p A )^2 F^* M^* + (\p A^* )^2 F M  - \frac{1}{3} M F^2 F^*  - \frac{1}{3} M (F^{*})^2 F \bigg)
\nn \\
\eea
and
\bea
\overbar{\La}_{3,\mathrm{II}} &=&
- \frac{1}{32} \int d^2 \Theta \, 2 \m{E} \, (\oD^2 - 8 R )\bigg( \frac{1}{8}c_2 + \frac{1}{3}c_3 - \frac{1}{3} c_4
+\frac{1}{\sqrt{2}}(\frac{2}{3}c_3 - \frac{4}{3} c_4) (\Phi + \Phi^\dagger)\bigg) \D \Phi \D \Phi \oD \Phi^\dagger \oD \Phi^\dagger + \mathrm{h.c.}
\nn \\
\nn \\
&=&
\bigg( \frac{1}{8}c_2 + \frac{1}{3}c_3 - \frac{1}{3} c_4
+\frac{1}{\sqrt{2}}(\frac{2}{3}c_3 - \frac{4}{3} c_4) (A + A^*)\bigg) 
\bigg(
4(\p A)^2 (\p A^*)^2 - 8 F F^*\p A \cdot \p A^*  + 4 (F F^*)^2
\bigg) \ .
\nn \\
\eea
Then
\bea
\overbar{\La}_3^{\mathrm{SUGRA}}  = \overbar{\La}_{3,\mathrm{I}} + \overbar{\La}_{3,\mathrm{II}} \, .
\eea

Ignoring, for the time being the contribution of $\overbar{\La}_4^{\mathrm{SUGRA}}$, let us take the $N=1$ supergravity Lagrangian for the worldvolume action of a probe brane in heterotic M-theory to be
\bea
\overbar{\La} &=& \overbar{\La}_1^{\mathrm{SUGRA}} + \overbar{\La}_2^{\mathrm{SUGRA}}  +\overbar{\La}_3^{\mathrm{SUGRA}} .
\label{L-total-sugra}
\eea
To ensure that this Lagrangian has the appropriate non-linear sigma model kinetic energy, that is, so that gravity is canonically normalized, one must perform a Weyl rescaling of the vielbein 
\bea
e_\mu^{~a} \rightarrow e_\mu^{~a} e^{\frac{1}{6} \Km} \ .
\eea
This induces transformations on the Ricci scalar $\mathcal{R}$ and on all covariant derivatives and Christoffel symbols in the component field Lagrangian. To proceed, one must now eliminate the auxiliary fields $M$ and $b_\mu$ using their equations of motion. The procedure is straightforward but tedious, and the essential steps were outlined in \cite{Deen:2017jqv}, Therefore, in this paper, we simply present the results. For compactness, we use the notation
\bea
&&D_A = \frac{\p}{\p A} + \frac{\p K}{\p A} \, , \qquad \bar{D}_{A^*} = \frac{\p}{\p A^*} + \frac{\p K}{\p A^*} , \qquad K_{, A} = \frac{\p K}{\p A}  \, , \qquad K_{, A^*} = \frac{\p K}{\p A^*}
\nn \\
&&\mu_1 = - \frac{1}{64} \bigg(  - \frac{1}{\sqrt{2}}c_3 - \frac{ \sqrt{2}}{3} c_4 \bigg) \, , ~~  \lambda_1 =  \frac{c_4}{32} , ~~ \mu_2 =   -\frac{1}{32} \bigg( \frac{1}{8} c_2 + \frac{1}{3}c_3 - \frac{1}{3} c_4 \bigg)  \, , ~~  \lambda_2 =  -\frac{1}{32 \sqrt{2}}  \big(\frac{2}{3}c_3 - \frac{4}{3}c_4 \big)
\nn \\
\nn \\
&&f(A, A^*) =  \frac{64}{3} \bigg( \mu_1 + \lambda_1(A + A^*)\bigg) \, \ .
\eea
Using this notation, we find that, after Weyl rescaling, the auxiliary field $b_\mu$ is given by
\bea
b_\mu = - \frac{3}{2} \bigg( j_\mu - \frac{1}{4\M}  \big( \mu_1 + \lambda_1(A + A^*)\big) h_\mu \bigg)
\eea
where
\bea
j_\mu &=& -\frac{i}{M_P^2} \bigg( K_{, A } \p_\mu A - K_{, A^* } \p_\mu A^*  \bigg)
\nn \\
h_\mu &=& i \bigg( 
\p_\mu A \big( \frac{512}{3}(\p A)^2 + \frac{128}{3} e^{\frac{1}{3} \x} F F^* \big) 
- \p_\mu A^* \big( \frac{512}{3}(\p A^*)^2 + \frac{128}{3} e^{\frac{1}{3} \x} F F^* \big)
\bigg) \ .
\label{jmu-hmu}
\eea
To remove the auxiliary field $M$, it is conventional to perform the following redefinition to another complex scalar $N$ defined by
\bea
M = N - \frac{1}{\M} \frac{\p K}{\p A^*} F^* \ .  
\eea
This, of course, leads to additional terms in \eqref{L-total-sugra} which depend on $F$ and $A$ alone. Solving for $N$, we find that
\bea
N &=& \frac{3}{\M}e^{-\frac{1}{3} \x} 
\bigg( - e^{\frac{2}{3} \x} W 
+ f e^{\frac{1}{3} \x} (\p A)^2 F^*
+ 3 f e^{\frac{2}{3} \x} (F^*)^2 F
\bigg) \ .
\eea
\noindent Inserting these results back into Lagrangian \eqref{L-total-sugra}, gives 
\bea
\frac{\bar{\m{L}}'}{e} &=&
-\tfrac{1}{2} M_P^2 \m{R} - K_{, A A^*} \p A \cdot \p A^* + e^{\frac{1}{3} \x} K_{, A A^*}  F F^* 
+ e^{\frac{2}{3} \x} (D_A W F + \bar{D}_{A^*} W^* F^* ) + \frac{3}{M_P^2} e^{\x} |W|^2 
\nn  \\
&&
- \frac{1}{4} \bigg( \mu_1 + \lambda_1(A + A^*)\bigg)
\bigg( 
16  (\p A)^2 \big( 16  \nabla^2 A^* 
+ 32  \p_\mu e^{ \frac{1}{6} \x} \p^\mu A^* \big)
\nn \\
&& \qquad \qquad \qquad \qquad \quad
+
16  (\p A^*)^2 \big( 16  \nabla^2A 
+ 32 \p_\mu e^{ \frac{1}{6} \x} \p^\mu A \big)
\bigg)
\nn \\ &&
- 128 e^{\frac{1}{3} \x} \bigg( \mu_1 + \lambda_1(A + A^*)\bigg) \bigg(
F^* \nabla F \cdot \nabla A + F \nabla F^* \cdot \nabla A^*
\bigg)
\nn \\ &&
+
\bigg( \mu_2 + \lambda_2(A + A^*)\bigg)
\bigg( -128 (\p A)^2 (\p A^*)^2 + 256 e^{\frac{1}{3} \x }\p A \cdot \p A^* F F^* - 128 e^{\frac{2}{3} \x } (F F^*)^2   \bigg)
\nn \\ &&
+ \lambda_1 \bigg( 128 e^{\frac{2}{3} \x } (FF^*)^2 - 256 e^{\frac{1}{3}\x}  \p A \cdot \p A^* F F^*- 64 e^{\frac{1}{3} \K} \big( (\p A)^2 + (\p A^*)^2 \big) FF^*\bigg)
\nn \\ &&
-\frac{f}{M_P^2} e^{\frac{1}{3} \x} FF^* \big( K_{, A} (\p A)^2 + K_{, A^*} (\p A^*)^2 \big)
-3\frac{f}{M_P^2} e^{\frac{2}{3} \x} (FF^*)^2 \big( K_{, A} + K_{, A^*} \big)
\nn \\ &&
+ \frac{3}{8}  \bigg( \mu_1 + \lambda_1(A + A^*)\bigg) j_\mu h^\mu
- \frac{3}{64 M_P^2}  \bigg( \mu_1 + \lambda_1(A + A^*)\bigg)^2 h_\mu h^\mu
\nn \\ &&
+ \frac{3}{M_P^2} \bigg(
 f^2 e^{\frac{2}{3} \x} (\p A)^2 (\p A^*)^2 F F^*
+ 9f^2 e^{\frac{4}{3} \x} (F F^*)^3
+ 3f^2e^{\x} (\p A)^2 (F F^*)^2  
\nn \\ &&
\qquad \qquad
+ 3f^2 e^{\x} (\p A^*)^2 (F F^*)^2 
-f e^{\x} \big(  W^* (\p A)^2 F^* + W (\p A^*)^2 F^*\big)
\nn \\ &&
\qquad \qquad
-3f e^{\frac{4}{3}\x} \big(  W^* (F^*)^2 F^* + W (F)^2 F^*\big)
\bigg) \, ,
\nn \\ \label{L-sugra}
\eea
where $j_\mu$, $h_\mu$ are given in \eqref{jmu-hmu}. The prime on $\bar{\m{L}}'$ indicates that both Weyl rescaling and the elimination of the supergravity auxiliary fields have been performed.

An important check on this result is the following. Taking the limit in which $M_P^2 \rightarrow \infty$, and $g_{\mu \nu} \rightarrow \eta_{\mu \nu}$, we find that 
\bea
\overbar{\La'} &=&
\frac{\p W}{\p A} F + \frac{\p W^*}{\p A^*} F^*
- \frac{\p^2 K}{\p A \p A^*} (\p A \cdot \p A^* - F F^*)
\nn \\
&+&
\bigg( -\frac{1}{\sqrt{2}} c_3 - \frac{\sqrt{2}}{3}c_4 - 2c_4(A + A^*)\bigg)
\bigg(
(\p A)^2 ( \Box  A^*) 
+
(\p A^*)^2 ( \Box A )
+ 2 F^* \p F \cdot \p A + 2 F \p F^* \cdot \p A^* 
\bigg)
\nn \\
&+&2c_4
\bigg( 2(F F^*)^2-  4 F F^* \p A \cdot \p A^*  -  F F^* \big( (\p A)^2 + (\p A^*)^2 \big)   \bigg)
\nn \\
&+&
\bigg( \frac{1}{8}c_2 + \frac{1}{3}c_3 - \frac{1}{3} c_4
+\frac{1}{\sqrt{2}}(\frac{2}{3}c_3 - \frac{4}{3} c_4) (A + A^*)\bigg) 
\bigg(
4(\p A)^2 (\p A^*)^2 - 8 F F^*\p A \cdot \p A^*  + 4 (F F^*)^2
\bigg) \ .
\nn \\
\eea
After an integration by parts, this is {\it precisely} the sum of the flat superspace Lagrangians presented in \eqref{L1-susy}, \eqref{L2-susy} and \eqref{L3-susy}--as it must be.

\subsection{$\overbar{\La}_4^{\mathrm{SUGRA}}$}

The $N=1$ supergravity extension of $\bar{\La}_4$ is given by 
\bea
\overbar{\La}_4^{\mathrm{SUGRA}} = \overbar{\La}_{4,\mathrm{I}}  + \overbar{\La}_{4,\mathrm{II}}  + \overbar{\La}_{4,\mathrm{II}} 
+\overbar{\La}_{4,\mathrm{IV}}  + \overbar{\La}_{4,\mathrm{V}}\, , 
\eea
where
\bea
\overbar{\La}_{4,\mathrm{I}} &=& \frac{-1}{256} \int d^2 \Theta \, 2 \m{E} \, (\oD^2 - 8 R )
\bigg(\frac{1}{4}c_4 + \frac{1}{6\sqrt{2}}(\Phi + \Phi^\dagger) \bigg) 
\{ \D, \oD \} (\D \Phi \D \Phi ) \{ \D, \oD \} (\oD\Phi^\dagger \oD \Phi^\dagger)\
+ \mathrm{h.c.} 
\nn \\
\nn \\
\overbar{\La}_{4,\mathrm{II}} &=& \frac{-1}{512} \int d^2 \Theta \, 2 \m{E} \, (\oD^2 - 8 R )
\bigg(c_4 + \frac{2}{3\sqrt{2}}c_4(\Phi + \Phi^\dagger) \bigg)  \{\D, \oD \} (\Phi + \Phi^\dagger)  \{\D, \oD \} (\D \Phi \D \Phi ) \oD^2 \Phi^\dagger 
+ \mathrm{h.c.}
\nn\\
\nn \\
\overbar{\La}_{4,\mathrm{III}} &=& \frac{-1}{256\sqrt{2}}\int d^2 \Theta \, 2 \m{E} \, (\oD^2 - 8 R )
\bigg(\frac{19}{6}c_{4} + \frac{76}{9 \sqrt{2}}c_{4} (\Phi + \Phi^\dagger) \bigg) 
\D \Phi \D \Phi \oD \Phi^\dagger \oD \Phi^\dagger \{\D, \oD \}  \{\D, \oD \} (\Phi + \Phi^\dagger) 
\nn \\
&&
+ \mathrm{h.c.}
\nn \\
\nn \\
\overbar{\La}_{4,\mathrm{IV}} &=& \frac{1}{512 \sqrt{2}}\int d^2 \Theta \, 2 \m{E} \, (\oD^2 - 8 R )
\bigg( -c_3 -\frac{11}{3}c_4 + \frac{1}{\sqrt{2}}(-\frac{2}{3}c_2 -\frac{88}{9}c_4)(\Phi + \Phi^\dagger) \bigg) 
\nn \\
&& \qquad \qquad \qquad \qquad \qquad \qquad 
\times~ \{ \D, \oD \} \D \Phi \D \Phi \oD \Phi \oD \Phi^\dagger \{ \D, \oD \} \Phi + \mathrm{h.c.}
\nn \\
\nn \\
\overbar{\La}_{4,\mathrm{V}} &=& -\frac{1}{128}\int d^2 \Theta \, 2 \m{E} \, (\oD^2 - 8 R )
\bigg( -\frac{1}{16} -\frac{1}{3}c_3 - \frac{9}{4}c_4 + \frac{1}{\sqrt{2}} (-\frac{1}{24} c_2 - \frac{8}{9}c_3 -\frac{21}{2}c_4)(\Phi + \Phi^\dagger) \bigg) 
\nn \\
&& \qquad \qquad \qquad \qquad \qquad \qquad
\times~ \D \Phi \D \Phi \oD \Phi^\dagger \oD \Phi^\dagger 
\{ \D, \oD \} \Phi \{ \D, \oD \} \Phi^\dagger + \mathrm{h.c.}
\nn \\
\eea
When expressed in components fields, $\overbar{\La}_4^{\mathrm{SUGRA}}$ will give rise to the appropriate higher derivatives of the complex scalar $A$, as well as those terms in \eqref{L4-1}-\eqref{L4-5} involving the auxiliary field $F$ and its derivatives. As in $\overbar{\La}_1^{\mathrm{SUGRA}}$, $\overbar{\La}_2^{\mathrm{SUGRA}}$ and $\overbar{\La}_3^{\mathrm{SUGRA}}$, we also find terms involving the auxiliary fields of supergravity. Now, however, there arise terms which are cubic or higher order in $M$, as well as terms involving {\it derivatives} of $b_\mu$ and $M$. We note that such terms were also present in the supergravity extension $\overbar{\La}_4^{\mathrm{SUGRA}}$ in the conformal Galileon case  discussed in \cite{Deen:2017jqv}. A complete analysis of these higher polynomial and derivative terms involving the supergravity auxiliary fields, both in the conformal and heterotic Galileon cases, will be discussed in detail elsewhere. However, this is not necessary in this paper, as we will see below.


\section{The Cosmological Limit}


Recall that the heterotic Galileons are derived in the limit where the four-dimensional momenta and the auxiliary field $F$ are all small compared the mass $\alpha$ associated with the curvature of the fifth-dimension. To continue in the gravitational case, it is extremely useful to work in a limit in which the four-dimensional spacetime curvature scalar $\mathcal{R}$ is restricted to be small compared to $\alpha^{2}$.  
That is, 
\bea
{\mathcal{R}}  \ll  \alpha^{2} < M_{CY}^{2} < M_P^2~\, . 
\label{pink1}
\eea
This scenario, which we referred to as the ``cosmological limit" in \cite{Deen:2017jqv}, allows one to drop the majority of terms appearing in the supergravity extended Lagrangian. 

Let us first consider the worldvolume Lagrangian \eqref{L-sugra}, constructed from $\overbar{\La}_1^{\mathrm{SUGRA}}$, $\overbar{\La}_2^{\mathrm{SUGRA}}$ and $\overbar{\La}_3^{\mathrm{SUGRA}}$ only. Taking this limit in \eqref{L-sugra}, the ``cosmological'' supergravity Lagrangian is found to be
\bea
\frac{\overbar{\La}^{\rm cosmo}_{1+2+3}}{e}&=&
- \frac{1}{2} \M \mathcal{R}
+ \frac{\p W}{\p A} F + \frac{\p W^*}{\p A^*} F^*
- \frac{\p^2 K}{\p A \p A^*} (\nabla A \cdot \nabla A^* - F F^*)
\nn \\
&+&
\bigg( -\frac{1}{\sqrt{2}} c_3 - \frac{\sqrt{2}}{3}c_4 - 2c_4(A + A^*)\bigg)
\bigg(
(\p A)^2 ( \nabla^2  A^*) 
+
(\p A^*)^2 ( \nabla^2 A )
\nn \\
&& \qquad \qquad 
+ 2 F^* \p F \cdot \p A + 2 F \p F^* \cdot \p A^* 
\bigg)
\nn \\
&+&2c_4
\bigg( 2(F F^*)^2 - 4 F F^* \nabla A \cdot \nabla A^*   -  F F^*\big( (\nabla A)^2 + (\nabla A^*)^2 \big) \bigg)
\nn \\
&+&
\bigg( \frac{1}{8}c_2 + \frac{1}{3}c_3 - \frac{1}{3} c_4
+\frac{1}{\sqrt{2}}(\frac{2}{3}c_3 - \frac{4}{3} c_4) (A + A^*)\bigg) 
\bigg(
4(\nabla A)^2 (\nabla A^*)^2 - 8 F F^*\nabla A \cdot \nabla A^*  
\nn \\
&& \qquad \qquad 
+ 4 (F F^*)^2
\bigg)  \ .
\nn \\
\label{Lcosmo-1}
\eea
We can now extend these results to include the ``cosmological'' terms from $\overbar{\La}_4^{\mathrm{SUGRA}}$. One can show that, in this limit, all terms arising from the elimination of the supergravity auxiliary fields $M$ and $b_{\mu}$ must necessarily be suppressed by powers of $M_{P}$  and, hence, can be ignored. The relevant terms will consist of two parts. The first part--labelled as $4A$--is made up of the expressions given in \eqref{L4-1}-\eqref{L4-5} with the partial derivatives $\p$ replaced by $\nabla$ and the metric $\eta_{\mu \nu}$ replaced with $g_{\mu \nu}$. That is, we add to \eqref{Lcosmo-1} the following:
\bea
\frac{\overline{\La}^{\rm cosmo}_{4A}}{e}
&=&
\bigg( - \frac{1}{4} c_4 - \frac{1}{6\sqrt{2}} c_4 (A + A^*)\bigg)
\bigg(
4\nabla_\mu (\nabla A)^2 \nabla^\mu (\nabla A^*)^2 
-8 \nabla_\mu (F A_{,\nu}) \nabla^\mu (F^* A^{*, \nu})
\nn \\
&&
\qquad \qquad 
+ 16 F F^* \nabla F \cdot \nabla F^*
\bigg)
\nn \\
%
&+&
\bigg( c_4 +  \frac{2}{3\sqrt{2}} c_4 (A + A^*) \bigg)
\bigg(
(A + A^*)^{, \mu} \big( \nabla_\mu (\nabla A)^2 \nabla^2 A +  \nabla_\mu (\nabla A^*)^2 \nabla^2 A^*\big)
\nn \\
&&
\qquad \qquad 
- (A + A^*)^{, \mu} \big( \nabla_\mu (F A_{, \nu}) F^{*, \nu} + \nabla_\mu (F^* A_{, \nu}^*) F^{, \nu}  \big)
\nn \\
&&
\qquad \qquad 
-  (A + A^*)^{, \mu} \big( \nabla_\mu (F \nabla^2 A - \nabla F \cdot \nabla A) F^*
+ \nabla_\mu (F^* \nabla^2 A^* - \nabla F^* \cdot \nabla A^*) F \big)
\nn \\
&&
\qquad \qquad 
- \nabla_\mu \nabla_\nu (A - A^*) \big( \nabla^\mu (F A^{, \nu}) F^* - \nabla^\mu (F^* A^{*, \nu}) F \big)
- 4 F F^* \nabla F \cdot \nabla F^*
\bigg)
\nn \\
&-&\frac{32}{3\sqrt{2}} c_4 (A - A^*)^{, \nu} (A + A^*)_{, \mu} 
\big( \nabla^\mu (F A_\nu) F^* -  \nabla^\mu (F^* A_\nu^*) F \big)
\nn \\
&-& \frac{32}{3\sqrt{2}} c_4 F F^*  (A + A^*)_{, \mu}
\big(  \nabla^\mu (\nabla A)^2 + \nabla^\mu (\nabla A^*)^2  \big)
\nn \\
&+& \frac{64}{3\sqrt{2}} c_4 \bigg( F (F^*)^2 (A + A^*)_{, \mu} F^{, \mu}  +  F^* F^2 (A + A^*)_{, \mu} F^{*, \mu}\bigg)
\nn \\
\nn \\
&+&
2\sqrt{2} \bigg( -\frac{19}{6} c_4 -\frac{76}{9\sqrt{2}}c_4 ( A + A^*)\bigg)
\bigg( (\nabla A )^2(\nabla A^*)^2 - 2 F F^* \nabla A \cdot \nabla A^* + (F F^*)^2\bigg) \nabla^2 (A + A^*) 
\nn \\
&+&
\frac{1}{\sqrt{2}} \bigg( -c_3 - \frac{11}{3} c_4 + \frac{1}{\sqrt{2}}(- \frac{2}{3} c_3 - \frac{88}{9}c_4) (A + A^*) \bigg)
\qquad \qquad \qquad \qquad \qquad \qquad \qquad
\nn \\
&&
\bigg( 
\nabla_\mu \big( (\nabla A)^2 (\nabla A^*)^2 \big) 
-2 \nabla_\mu (F F^* \nabla A \cdot \nabla A^*) 
+ \nabla_\mu (F F^*)^2
\bigg)\big( A + A^*\big)^{, \mu}
\nn \\
&+&
\bigg( 
-\frac{1}{16} c_2 - \frac{1}{3} c_3 - \frac{9}{4} c_4 
+ \frac{1}{\sqrt{2}} (-\frac{1}{24} c_2 -\frac{8}{9}c_3 - \frac{21}{2} c_4) 
(A + A^*)
\bigg) \qquad \qquad \qquad 
\nn \\
&&
\bigg(
(\nabla A )^2(\nabla A^*)^2 - 2 F F^* \nabla A \cdot \nabla A^* + (F F^*)^2
\bigg) \nabla A \cdot \nabla A^*
\eea
One finds, however, that in addition to these terms, the presence of curvature leads to {\it two additional terms} arising from the supergravity extension of $\overbar{\La}_{4, \, \mathrm{2nd term}}^{\mathrm{SUSY}}$ in \eqref{L4-2} that are not suppressed in the cosmological limit. These constitute the second part of the $\overbar{\La}_4^{\mathrm{SUGRA}}$ contribution to the cosmological limit and are given by
\bea
\frac{\overline{\La}^{\rm cosmo}_{4B}}{e} &=&
\bigg( c_4 + \frac{2}{3\sqrt{2}}c_4(A + A^*)\bigg)
\bigg(
\frac{17}{4} \mathcal{R} F F^* \nabla^2 (A + A^*) - \frac{9}{8} F F^* \mathcal{R}_{\mu \nu} \nabla^\mu (A + A^*) \nabla^\nu (A + A^*)
\bigg) \, .
\nn \\
\eea

We conclude that in the cosmological limit defined by \eqref{pink1}, the diffeomorphically invariant four-dimensional $N=1$ supersymmetric Lagrangian is given by the sum
\be
\frac{\overline{\La}^{\rm cosmo}}{e} = \frac{\overbar{\La}^{\rm cosmo}_{1+2+3+4(A+B)}}{e} \ .
\label{pink2}
\ee
For completeness, the entire ``cosmological'' Lagrangian density is presented in Appendix B.

\section*{Acknowledgments}
R. Deen and B.A. Ovrut are supported in part by DOE contract No. $\mathrm{DE}$-$\rm{SC}0007901$. B. Ovrut would like to thank Anna Ijjas , Paul Steinhardt and other members of the PCTS working group ``Rethinking Cosmology'' for many helpful conversations. Ovrut would also like to thank his long term collaborator Jean-Luc Lehners for his joint work on higher-derivative supersymmetry and supergravitation. Finally, R. Deen is grateful to Anna Ijjas and Paul Steinhardt for many discussions and the Center for Particle Cosmology at the University of Pennsylvania for their support. Figures 1-3 in the text were created using \textit{Matplotlib}, \cite{Hunter:2007}.

\section*{Appendix A ~ The Total $N=1$ Supersymmetric Lagrangian}
The flat $N=1$ superspace Lagrangian density arising from the first four heterotic Galileons is given by
\bea
\overbar{\La}^{\mathrm{SUSY}}_{1+2+3+4} 
&=& \beta_{1}(F+F^{*})+ \sqrt{2}\beta_{2}(F+F^{*})\hpi +i\sqrt{2}\beta_{2}(F-F^{*})\chi 
\nn \\
&+&\bigg(- \frac{1}{2} c_2 -  c_3  - \frac{2}{3}c_4+ ( \frac{1}{3}c_2 - \frac{4}{3}c_3 - \frac{20}{9} c_4) \hpi\bigg) \bigg( (\frac{\p \hpi}{\alpha})^2 + (\frac{\p \chi}{\alpha})^2 - 2\frac{F F^*}{\alpha^2}
\bigg) 
\nn \\
&+&
\bigg( - \frac{1}{2}c_3 - c_4 - 2 c_4 \hpi \bigg)
 \bigg(
(\p \hpi)^2 \Box \hpi + (\p \chi)^2 \Box \hpi + 2 \p \hpi \cdot \p \chi \Box \chi
- 2F F^* \Box \hpi
\nn \\ &&
\qquad \qquad 
+ 2iF^* \p F \cdot \p \chi -  2i F \p F^* \cdot \p \chi
\bigg)
+4 c_4 
\bigg( 
 (FF^*)^2 + \frac{1}{2} FF^* (\p \hpi)^2 + \frac{1}{2}FF^* (\p \chi)^2  
\bigg)
\nn \\
&+&  
\bigg( \frac{1}{8} c_2 + \frac{1}{3}c_3 - \frac{1}{3} c_4 + \big(\frac{2}{3}c_3 - \frac{4}{3}c_4 \big) \hpi \bigg)
\bigg((\p \hpi)^4 + (\p \chi)^4 - 2 (\p \hpi )^2 (\p \chi)^2 + 4(\p \hpi \cdot \p \chi)^2  
\nn \\
&&\qquad \qquad \qquad 
 -   4 FF^* (\p \hpi)^2 - 4 FF^* (\p \chi)^2    +  4 (FF^*)^2 \bigg) 
\nn \\
&+&
\bigg( - \frac{1}{4} c_4 - \frac{1}{6} c_4  \hpi \bigg)
\bigg(
\p_\mu (\p \hpi)^2 \p^\mu (\p \hpi)^2 + \p_\mu (\p \chi)^2 \p^\mu (\p \chi)^2 - 2 \p_\mu (\p \hpi)^2 \p^\mu (\p \chi)^2 
\nn \\
&&
\qquad  \qquad 
+ 4 \p_\mu (\p \hpi \cdot \p \chi )\p^\mu (\p \hpi \cdot \p \chi) 
- 4 \p_\mu (F \hpi_{,\nu}) \p^\mu (F^* \hpi^{,\nu})
\nn \\
&&
\qquad \qquad 
- 4 \p_\mu (F \chi_{,\nu}) \p^\mu (F^* \chi^{,\nu})
+ 16 F F^* \p F \cdot \p F^*
\bigg)
\nn \\
&+&\bigg( c_4 +  \frac{2}{3} c_4 \hpi \bigg)
\bigg(
\Box \hpi \p^\mu (\p \hpi)^2 \hpi_{, \mu} - \Box \hpi \p^{\mu}(\p \chi)^2 \hpi_{, \mu} - 2\Box \chi \p^\mu (\p \hpi \cdot \p \chi) \hpi_{, \mu}
\nn \\
&&
\qquad 
- \hpi^{, \mu } \p_{\mu} (F \hpi_{, \nu}) F^{*, \nu} - \hpi^{, \mu } \p_{\mu} (F^* \hpi_{, \nu}) F^{, \nu}
- i \hpi^{, \mu } \p_{\mu} (F \chi_{, \nu}) F^{*, \nu} + i \hpi^{, \mu } \p_{\mu} (F^* \chi_{, \nu}) F^{, \nu}
\nn \\
&&
\qquad  
- \hpi^{, \mu } \p_\mu ( F \Box \hpi)F^* - \hpi^{, \mu} \p_{\mu} (F^* \Box \hpi)F
- \hpi^{, \mu } \p_\mu ( \p F \cdot \p \hpi) F^* - \hpi^{, \mu} \p_{\mu} (\p F^* \cdot \p \hpi)F
\nn \\
&&
\qquad 
- i \hpi^{, \mu } \p_\mu ( F \Box \chi)F^* + i \hpi^{, \mu} \p_{\mu} (F^* \Box \chi)F
+ i\hpi^{, \mu } \p_\mu ( \p F \cdot \p \chi) F^* - i \hpi^{, \mu} \p_{\mu} (\p F^* \cdot \p \chi)F
\nn \\
&&
\qquad 
- i \chi_{, \mu \nu} \p^{\mu}(F \hpi^{, \nu})F^* + i  \chi_{, \mu \nu} \p^{\mu}(F^* \hpi^{, \nu})F
\qquad  
+ \chi_{, \mu \nu} \p^{\mu}(F \chi^{, \nu})F^* +   \chi_{, \mu \nu} \p^{\mu}(F^* \chi^{, \nu})F
\nn \\
&&
\qquad
- 4 F F^* \p F \cdot \p F^*
\bigg)
\nn \\
&+& \frac{32}{3} c_4 \bigg(
i \hpi^{, \mu} \p_{\mu} (F \hpi_{, \nu}) \chi^{, \nu} F^* - i \hpi^{, \mu} \p_{\mu} (F^* \hpi_{, \nu}) \chi^{, \nu} F
-  \hpi^{, \mu} \p_{\mu} (F \chi_{, \nu}) \chi^{, \nu} F^* - \hpi^{, \mu} \p_{\mu} (F^* \chi_{, \nu}) \chi^{, \nu} F
\bigg)
\nn 
\eea
\bea
&-& \frac{32}{3} c_4 \bigg(
F F^* \hpi^{, \mu} \p_{\mu} (\p \hpi)^2 + F F^* \hpi^{, \mu} \p_\mu (\p \chi)^2 
\bigg)
+\frac{64}{3} c_4 \bigg(
F (F^*)2 \p F \cdot \p \hpi + F^* F^2 \p F^* \cdot \p \hpi
\bigg)
\nn \\
&+&
 \bigg( - \frac{19}{6} c_4 - \frac{76}{9}c_4 \hpi \bigg) 
 \bigg(
 (\p \hpi)^4 \Box \hpi + (\p \chi)^4 \Box \hpi - 2 (\p \hpi)^2 (\p \chi)^2 \Box \hpi
 + (\p \hpi \cdot \p \chi)^2 \Box \hpi 
 \nn \\
 &&
 \qquad \qquad 
 - 2 F F^* ( (\p \hpi)^2 + (\p \chi)^2 ) \Box \hpi
 + 4 (F F^* )^2 \Box \hpi
 \bigg)
\nn \\
&+&
\bigg( -c_3 - \frac{11}{3} c_4 + (- \frac{2}{3} c_3 - \frac{88}{9}c_4) \hpi  \bigg)
\bigg(
(\p \hpi)^2 \hpi_{,\mu} \hpi_{\mu \nu} \hpi^{, \nu}
- (\p \hpi)^2 \hpi_{,\mu} \hpi_{\mu \nu} \hpi^{, \nu}
\nn \\ 
&&
\qquad \qquad 
- (\p \hpi)^2 \hpi_{,\mu} \chi_{\mu \nu} \chi^{, \nu}
+ (\p \chi)^2 \hpi_{,\mu} \chi_{\mu \nu} \chi^{, \nu}
+ 4 \p_\mu \big( \p \hpi \cdot \p \chi \big)^2 \hpi^{, \mu}
- \p_\mu (F F^* (\p \hpi)^2 ) \hpi^{, \mu} 
\nn \\ 
&&
\qquad \qquad 
- \p_\mu (F F^* (\p \chi)^2 ) \hpi^{, \mu} 
+ \p_\mu(F F^*) \hpi^{, \mu}
\bigg)
\nn \\
&+&
\bigg(
-\frac{1}{16} c_2 - \frac{1}{3} c_3 - \frac{9}{4} c_4 
+ (-\frac{1}{24} c_2 -\frac{8}{9}c_3 - \frac{21}{2} c_4)  \hpi
\bigg)
\bigg(
(\p \hpi)^2 + (\p \chi)^2
\bigg)
\nn \\
&&
\cdot
\bigg( 
(\p \hpi)^4 + (\p \chi)^4 - 2 (\p \hpi)^2 (\p \chi)^2 + 4 (\p \hpi \cdot \p \chi)^2 
- 4 F F^* \big( (\p \hpi)^2 - (\p \chi)^2\big)
+ 4 (F F^*)^2
\bigg) \, ,
\nn \\ 
\eea
where we have dropped all terms containing the chiral fermion $\psi$.

\begin{figure}[!hb]
\begin{center}
\includegraphics[scale=0.3]{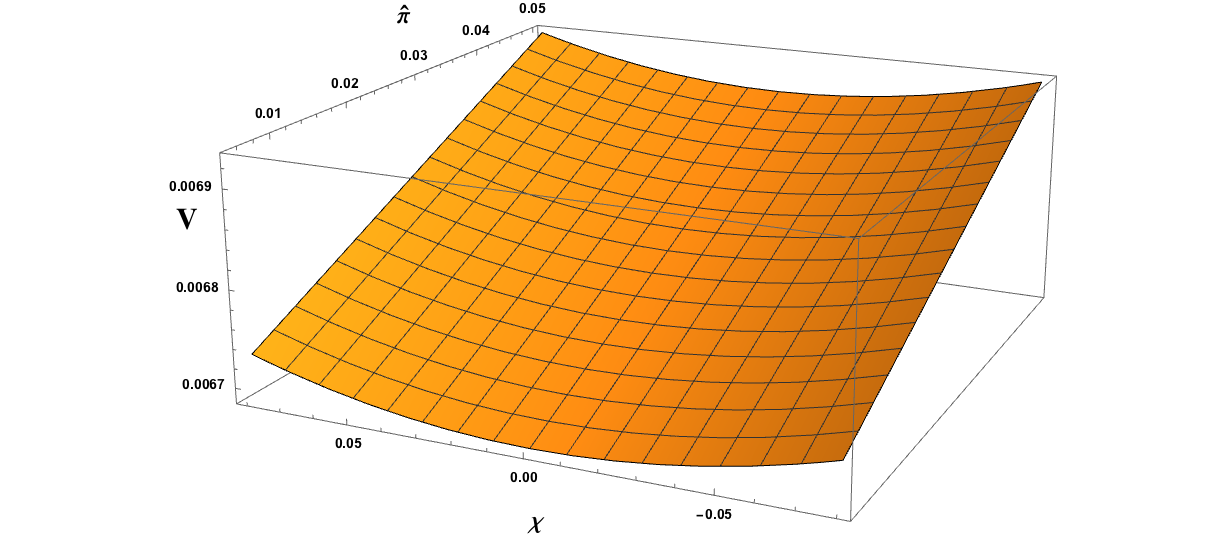}
\caption{\small{Plot of the potential energy $V$ when the inequalities given in the Summary of Constraints above are satisfied. Here,  $(c_1, c_2, c_3, c_4) = (-1.00,  0.60, -0.22,  0.46)$. We note the positive slope in the $\hpi$ direction and the local minimum of the potential at $\chi=0$ for any fixed value of $\hpi$, as we required.}}
\label{potential-plot}
\end{center}
\end{figure}

As discussed in Section 4, the $F$ term can be evaluated using a perturbative expansion. A good approximation to $F$ is given by the lowest order term
\be
F^{(0)} =
 -\frac{\beta_1}{\gamma} + \sqrt{2} \bigg( 2 \frac{\beta_1}{\gamma} \bigg(\frac{\delta}{\gamma}\bigg) - \frac{\beta_2}{ \gamma} \bigg) \hpi
+ i \sqrt{2} \frac{\beta_2}{\gamma} \chi
- i4\frac{\beta_2}{\gamma} \bigg(\frac{\delta}{\gamma}\bigg)  \hpi \chi \, ,
\nn \\
\label{car1}
\ee
where $\gamma, \delta$ are defined in equation \eqref{gamma-delta}, $\beta_1$, $\beta_{2}$ are real coefficients of the superpotential defined in \eqref{W-1} and the coefficients $c_{i}$, $i=1,2,3,4$ satisfy the constraints in Subsection 4.7. In Figure \ref{potential-plot}, we plot the leading order potential in equation \eqref{V}, for a specific case in which the constraints are satisfied.

\section*{Appendix B ~ The Total Cosmological Lagrangian}
The Lagrangian density arising from the first four $N=1$ supergravitational  heterotic Galileons in the ``cosmological'' limit is given, in component fields, by
\bea
\frac{\overline{\La}^{\rm cosmo}_{1+2+3+4}}{e} &=& - \frac{1}{2} \M \mathcal{R}
+ \frac{\p W}{\p A} F + \frac{\p W^*}{\p A^*} F^*
- \frac{\p^2 K}{\p A \p A^*} (\nabla A \cdot \nabla A^* - F F^*)
\nn \\
&+&
\bigg( -\frac{1}{\sqrt{2}} c_3 - \frac{\sqrt{2}}{3}c_4 - 2c_4(A + A^*)\bigg)
\bigg(
(\p A)^2 ( \nabla^2  A^*) 
+
(\p A^*)^2 ( \nabla^2 A )
\nn \\
&& \qquad \qquad 
+ 2 F^* \p F \cdot \p A + 2 F \p F^* \cdot \p A^* 
\bigg)
\nn \\
&+&2c_4
\bigg( 2(F F^*)^2 - 4 F F^* \nabla A \cdot \nabla A^*   -  F F^*\big( (\nabla A)^2 + (\nabla A^*)^2 \big) \bigg)
\nn \\
&+&
\bigg( \frac{1}{8}c_2 + \frac{1}{3}c_3 - \frac{1}{3} c_4
+\frac{1}{\sqrt{2}}(\frac{2}{3}c_3 - \frac{4}{3} c_4) (A + A^*)\bigg) 
\bigg(
4(\nabla A)^2 (\nabla A^*)^2 - 8 F F^*\nabla A \cdot \nabla A^*  + 4 (F F^*)^2
\bigg)  
\nn \\
&+&
\bigg( - \frac{1}{4} c_4 - \frac{1}{6\sqrt{2}} c_4 (A + A^*)\bigg)
\bigg(
4\nabla_\mu (\nabla A)^2 \nabla^\mu (\nabla A^*)^2 
-8 \nabla_\mu (F A_{,\nu}) \nabla^\mu (F^* A^{*, \nu}) \qquad \qquad
\nn \\
&&
\qquad \qquad
+ 16 F F^* \nabla F \cdot \nabla F^*
\bigg)
\nn \\
&+&
\bigg( c_4 +  \frac{2}{3\sqrt{2}} c_4 (A + A^*) \bigg)
\bigg(
(A + A^*)^{, \mu} \big( \nabla_\mu (\nabla A)^2 \nabla^2 A +  \nabla_\mu (\nabla A^*)^2 \nabla^2 A^*\big)
\nn \\
&&
\qquad \qquad 
- (A + A^*)^{, \mu} \big( \nabla_\mu (F A_{, \nu}) F^{*, \nu} + \nabla_\mu (F^* A_{, \nu}^*) F^{, \nu}  \big)
\nn \\
&&
\qquad \qquad 
-  (A + A^*)^{, \mu} \big( \nabla_\mu (F \nabla^2 A - \nabla F \cdot \nabla A) F^*
+ \nabla_\mu (F^* \nabla^2 A^* - \nabla F^* \cdot \nabla A^*) F \big)
\nn \\
&&
\qquad \qquad 
- \nabla_\mu \nabla_\nu (A - A^*) \big( \nabla^\mu (F A^{, \nu}) F^* - \nabla^\mu (F^* A^{*, \nu}) F \big)
- 4 F F^* \nabla F \cdot \nabla F^*
\bigg)
\nn \\
&-&\frac{32}{3\sqrt{2}} c_4 (A - A^*)^{, \nu} (A + A^*)_{, \mu} 
\big( \nabla^\mu (F A_\nu) F^* -  \nabla^\mu (F^* A_\nu^*) F \big)
\nn \\
&-& \frac{32}{3\sqrt{2}} c_4 F F^*  (A + A^*)_{, \mu}
\big(  \nabla^\mu (\nabla A)^2 + \nabla^\mu (\nabla A^*)^2  \big)
\nn \\
&+& \frac{64}{3\sqrt{2}} c_4 \bigg( F (F^*)^2 (A + A^*)_{, \mu} F^{, \mu}  +  F^* F^2 (A + A^*)_{, \mu} F^{*, \mu}\bigg)
\nn \\
\nn \\
&+&
2\sqrt{2} \bigg( -\frac{19}{6} c_4 -\frac{76}{9\sqrt{2}}c_4 ( A + A^*)\bigg)
\bigg( (\nabla A )^2(\nabla A^*)^2 - 2 F F^* \nabla A \cdot \nabla A^* + (F F^*)^2\bigg) \nabla^2 (A + A^*) 
\nn \\
&+&
\frac{1}{\sqrt{2}} \bigg( -c_3 - \frac{11}{3} c_4 + \frac{1}{\sqrt{2}}(- \frac{2}{3} c_3 - \frac{88}{9}c_4) (A + A^*) \bigg)
\nn \\
&&
\bigg( 
\nabla_\mu \big( (\nabla A)^2 (\nabla A^*)^2 \big) 
-2 \nabla_\mu (F F^* \nabla A \cdot \nabla A^*) 
+ \nabla_\mu (F F^*)^2
\bigg)\big( A + A^*\big)^{, \mu}
\nn \\
&+&
\bigg( 
-\frac{1}{16} c_2 - \frac{1}{3} c_3 - \frac{9}{4} c_4 
+ \frac{1}{\sqrt{2}} (-\frac{1}{24} c_2 -\frac{8}{9}c_3 - \frac{21}{2} c_4) 
(A + A^*)
\bigg) \qquad
\nn \\
&&
\quad \bigg(
(\nabla A )^2(\nabla A^*)^2 - 2 F F^* \nabla A \cdot \nabla A^* + (F F^*)^2
\bigg) \nabla A \cdot \nabla A^*
\nn \\
&+&
\bigg( c_4 + \frac{2}{3\sqrt{2}}c_4(A + A^*)\bigg)
\bigg(
\frac{17}{4} \mathcal{R} F F^* \nabla^2 (A + A^*) - \frac{9}{8} F F^* \mathcal{R}_{\mu \nu} \nabla^\mu (A + A^*) \nabla^\nu (A + A^*)
\bigg) \, ,
\label{eq-111}
\eea
where $A=\frac{1}{\sqrt{2}}(\hpi+\chi)$ and we have dropped all terms containing the chiral fermion $\psi$ and the gravitino $\psi^{\alpha}_{\mu}$. The Kahler potential $K$ and the superpotential $W$ are given by
\bea
K(A , A^\dagger) = \gamma A A^\dagger + \delta (A^2 A^\dagger + A (A^\dagger)^2) \, ,
\qquad
W(\Phi) = \beta_1 A + \beta_2 A^2 \, ,
\eea
where $\gamma, \delta$ are defined in equation \eqref{gamma-delta} and $\beta_1$, $\beta_{2}$ are real coefficients. We emphasize again that the Lagrangian density \eqref{eq-111} is derived in the ``cosmological" limit.  Hence, the equations of motion arising from it must be solved order by order in powers of $(\partial/\alpha)^2$.


\begin{thebibliography}{99}

\bibitem{Aad:2012tfa} 
  G.~Aad {\it et al.} [ATLAS Collaboration],
  ``Observation of a new particle in the search for the Standard Model Higgs boson with the ATLAS detector at the LHC,''
  Phys.\ Lett.\ B {\bf 716}, 1 (2012)
  doi:10.1016/j.physletb.2012.08.020
  [arXiv:1207.7214 [hep-ex]].

\bibitem{Chatrchyan:2012xdj} 
  S.~Chatrchyan {\it et al.} [CMS Collaboration],
  ``Observation of a new boson at a mass of 125 GeV with the CMS experiment at the LHC,''
  Phys.\ Lett.\ B {\bf 716}, 30 (2012)
  doi:10.1016/j.physletb.2012.08.021
  [arXiv:1207.7235 [hep-ex]].

\bibitem{Bezrukov:2007ep} 
  F.~L.~Bezrukov and M.~Shaposhnikov,
  ``The Standard Model Higgs boson as the inflaton,''
  Phys.\ Lett.\ B {\bf 659}, 703 (2008)
  doi:10.1016/j.physletb.2007.11.072
  [arXiv:0710.3755 [hep-th]].

\bibitem{Einhorn:2009bh} 
  M.~B.~Einhorn and D.~R.~T.~Jones,
  ``Inflation with Non-minimal Gravitational Couplings in Supergravity,''
  JHEP {\bf 1003}, 026 (2010)
  doi:10.1007/JHEP03(2010)026
  [arXiv:0912.2718 [hep-ph]].

\bibitem{Ferrara:2010yw} 
  S.~Ferrara, R.~Kallosh, A.~Linde, A.~Marrani and A.~Van Proeyen,
  ``Jordan Frame Supergravity and Inflation in NMSSM,''
  Phys.\ Rev.\ D {\bf 82}, 045003 (2010)
  doi:10.1103/PhysRevD.82.045003
  [arXiv:1004.0712 [hep-th]].

\bibitem{Deen:2016zfr} 
  R.~Deen, B.~A.~Ovrut and A.~Purves,
  ``Supersymmetric Sneutrino-Higgs Inflation,''
  Phys.\ Lett.\ B {\bf 762}, 441 (2016)
  doi:10.1016/j.physletb.2016.09.059
  [arXiv:1606.00431 [hep-ph]].
  
\bibitem{Khoury:2001wf} 
  J.~Khoury, B.~A.~Ovrut, P.~J.~Steinhardt and N.~Turok,
  ``The Ekpyrotic universe: Colliding branes and the origin of the hot big bang,''
  Phys.\ Rev.\ D {\bf 64}, 123522 (2001)
  doi:10.1103/PhysRevD.64.123522
  [hep-th/0103239].
  
\bibitem{Khoury:2001bz} 
  J.~Khoury, B.~A.~Ovrut, N.~Seiberg, P.~J.~Steinhardt and N.~Turok,
  ``From big crunch to big bang,''
  Phys.\ Rev.\ D {\bf 65}, 086007 (2002)
  doi:10.1103/PhysRevD.65.086007
  [hep-th/0108187]. 
  
\bibitem{Khoury:2001zk} 
  J.~Khoury, B.~A.~Ovrut, P.~J.~Steinhardt and N.~Turok,
  ``Density perturbations in the ekpyrotic scenario,''
  Phys.\ Rev.\ D {\bf 66}, 046005 (2002)
  doi:10.1103/PhysRevD.66.046005
  [hep-th/0109050].
  
\bibitem{Buchbinder:2007ad} 
  E.~I.~Buchbinder, J.~Khoury and B.~A.~Ovrut,
  ``New Ekpyrotic cosmology,''
  Phys.\ Rev.\ D {\bf 76}, 123503 (2007)
  doi:10.1103/PhysRevD.76.123503
  [hep-th/0702154].
  
\bibitem{Buchbinder:2007tw} 
  E.~I.~Buchbinder, J.~Khoury and B.~A.~Ovrut,
  ``On the initial conditions in new ekpyrotic cosmology,''
  JHEP {\bf 0711}, 076 (2007)
  doi:10.1088/1126-6708/2007/11/076
  [arXiv:0706.3903 [hep-th]].
  
\bibitem{Buchbinder:2007at} 
  E.~I.~Buchbinder, J.~Khoury and B.~A.~Ovrut,
  ``Non-Gaussianities in new ekpyrotic cosmology,''
  Phys.\ Rev.\ Lett.\  {\bf 100}, 171302 (2008)
  doi:10.1103/PhysRevLett.100.171302
  [arXiv:0710.5172 [hep-th]].
   
 \bibitem{Creminelli:2007aq} 
  P.~Creminelli and L.~Senatore,
  ``A Smooth bouncing cosmology with scale invariant spectrum,''
  JCAP {\bf 0711}, 010 (2007)
  doi:10.1088/1475-7516/2007/11/010
  [hep-th/0702165].
  
\bibitem{Easson:2011zy} 
  D.~A.~Easson, I.~Sawicki and A.~Vikman,
  ``G-Bounce,''
  JCAP {\bf 1111}, 021 (2011)
  doi:10.1088/1475-7516/2011/11/021
  [arXiv:1109.1047 [hep-th]].
  
\bibitem{Cai:2012va} 
  Y.~F.~Cai, D.~A.~Easson and R.~Brandenberger,
  ``Towards a Nonsingular Bouncing Cosmology,''
  JCAP {\bf 1208}, 020 (2012)
  doi:10.1088/1475-7516/2012/08/020
  [arXiv:1206.2382 [hep-th]].

\bibitem{Brandenberger:2012zb} 
  R.~H.~Brandenberger,
  ``The Matter Bounce Alternative to Inflationary Cosmology,''
  arXiv:1206.4196 [astro-ph.CO].
  
\bibitem{Brandenberger:2016vhg} 
  R.~Brandenberger and P.~Peter,
  ``Bouncing Cosmologies: Progress and Problems,''
  doi:10.1007/s10701-016-0057-0
  arXiv:1603.05834 [hep-th].
 
\bibitem{Koehn:2013upa} 
  M.~Koehn, J.~L.~Lehners and B.~A.~Ovrut,
  ``Cosmological super-bounce,''
  Phys.\ Rev.\ D {\bf 90}, no. 2, 025005 (2014)
  doi:10.1103/PhysRevD.90.025005
  [arXiv:1310.7577 [hep-th]].
  
\bibitem{Koehn:2015vvy} 
  M.~Koehn, J.~L.~Lehners and B.~Ovrut,
  ``Nonsingular bouncing cosmology: Consistency of the effective description,''
  Phys.\ Rev.\ D {\bf 93}, no. 10, 103501 (2016)
  doi:10.1103/PhysRevD.93.103501
  [arXiv:1512.03807 [hep-th]].
  
\bibitem{Ijjas:2016tpn} 
  A.~Ijjas and P.~J.~Steinhardt,
  ``Classically stable nonsingular cosmological bounces,''
  Phys.\ Rev.\ Lett.\  {\bf 117}, no. 12, 121304 (2016)
  doi:10.1103/PhysRevLett.117.121304
  [arXiv:1606.08880 [gr-qc]].
   
\bibitem{Ijjas:2016vtq} 
  A.~Ijjas and P.~J.~Steinhardt,
  ``Fully stable cosmological solutions with a non-singular classical bounce,''
  Phys.\ Lett.\ B {\bf 764}, 289 (2017)
  doi:10.1016/j.physletb.2016.11.047
  [arXiv:1609.01253 [gr-qc]].
   
     
\bibitem{Creminelli:2010ba} 
  P.~Creminelli, A.~Nicolis and E.~Trincherini,
  ``Galilean Genesis: An Alternative to inflation,''
  JCAP {\bf 1011}, 021 (2010)
  doi:10.1088/1475-7516/2010/11/021
  [arXiv:1007.0027 [hep-th]].
 
\bibitem{Hinterbichler:2012yn} 
  K.~Hinterbichler, A.~Joyce, J.~Khoury and G.~E.~J.~Miller,
  ``Dirac-Born-Infeld Genesis: An Improved Violation of the Null Energy Condition,''
  Phys.\ Rev.\ Lett.\  {\bf 110}, no. 24, 241303 (2013)
  doi:10.1103/PhysRevLett.110.241303
  [arXiv:1212.3607 [hep-th]].
   
\bibitem{Goon:2011qf} 
  G.~Goon, K.~Hinterbichler and M.~Trodden,
  ``Symmetries for Galileons and DBI scalars on curved space,''
  JCAP {\bf 1107}, 017 (2011)
  doi:10.1088/1475-7516/2011/07/017
  [arXiv:1103.5745 [hep-th]].
  
\bibitem{Goon:2011uw} 
  G.~Goon, K.~Hinterbichler and M.~Trodden,
  ``A New Class of Effective Field Theories from Embedded Branes,''
  Phys.\ Rev.\ Lett.\  {\bf 106}, 231102 (2011)
  doi:10.1103/PhysRevLett.106.231102
  [arXiv:1103.6029 [hep-th]].
   
\bibitem{Dvali:2000hr} 
  G.~R.~Dvali, G.~Gabadadze and M.~Porrati,
  ``4-D gravity on a brane in 5-D Minkowski space,''
  Phys.\ Lett.\ B {\bf 485}, 208 (2000)
  doi:10.1016/S0370-2693(00)00669-9
  [hep-th/0005016].
    
\bibitem{Nicolis:2008in} 
  A.~Nicolis, R.~Rattazzi and E.~Trincherini,
  ``The Galileon as a local modification of gravity,''
  Phys.\ Rev.\ D {\bf 79}, 064036 (2009)
  doi:10.1103/PhysRevD.79.064036
  [arXiv:0811.2197 [hep-th]].
 
\bibitem{deRham:2010eu} 
  C.~de Rham and A.~J.~Tolley,
  ``DBI and the Galileon reunited,''
  JCAP {\bf 1005}, 015 (2010)
  doi:10.1088/1475-7516/2010/05/015
  [arXiv:1003.5917 [hep-th]].
  

\bibitem{Lukas:1997fg} 
  A.~Lukas, B.~A.~Ovrut and D.~Waldram,
  ``On the four-dimensional effective action of strongly coupled heterotic string theory,''
  Nucl.\ Phys.\ B {\bf 532}, 43 (1998)
  doi:10.1016/S0550-3213(98)00463-5
  [hep-th/9710208].
  
\bibitem{Lukas:1998yy} 
  A.~Lukas, B.~A.~Ovrut, K.~S.~Stelle and D.~Waldram,
  ``The Universe as a domain wall,''
  Phys.\ Rev.\ D {\bf 59}, 086001 (1999)
  doi:10.1103/PhysRevD.59.086001
  [hep-th/9803235].
  
\bibitem{Lukas:1998tt} 
  A.~Lukas, B.~A.~Ovrut, K.~S.~Stelle and D.~Waldram,
  ``Heterotic M theory in five-dimensions,''
  Nucl.\ Phys.\ B {\bf 552}, 246 (1999)
  doi:10.1016/S0550-3213(99)00196-0
  [hep-th/9806051].
     
\bibitem{Ovrut:2012wn} 
  B.~A.~Ovrut and J.~Stokes,
  ``Heterotic Kink Solitons and their Worldvolume Action,''
  JHEP {\bf 1209}, 065 (2012)
  doi:10.1007/JHEP09(2012)065
  [arXiv:1205.4236 [hep-th]].
  
\bibitem{Deen:2017jqv} 
  R.~Deen and B.~Ovrut,
  ``Supergravitational Conformal Galileons,''
  arXiv:1705.06729 [hep-th].
   
\bibitem{Horava:1995qa} 
  P.~Horava and E.~Witten,
  ``Heterotic and type I string dynamics from eleven-dimensions,''
  Nucl.\ Phys.\ B {\bf 460}, 506 (1996)
  doi:10.1016/0550-3213(95)00621-4
  [hep-th/9510209].
  
\bibitem{Horava:1996ma} 
  P.~Horava and E.~Witten,
  ``Eleven-dimensional supergravity on a manifold with boundary,''
  Nucl.\ Phys.\ B {\bf 475}, 94 (1996)
  doi:10.1016/0550-3213(96)00308-2
  [hep-th/9603142].
  
\bibitem{Lukas:1998hk} 
  A.~Lukas, B.~A.~Ovrut and D.~Waldram,
  ``Nonstandard embedding and five-branes in heterotic M theory,''
  Phys.\ Rev.\ D {\bf 59}, 106005 (1999)
  doi:10.1103/PhysRevD.59.106005
  [hep-th/9808101].
  
\bibitem{Donagi:1999gc} 
  R.~Donagi, A.~Lukas, B.~A.~Ovrut and D.~Waldram,
  ``Holomorphic vector bundles and nonperturbative vacua in M theory,''
  JHEP {\bf 9906}, 034 (1999)
  doi:10.1088/1126-6708/1999/06/034
  [hep-th/9901009].
  
\bibitem{Donagi:2004ia}
  R.~Donagi, Y.~H.~He, B.~A.~Ovrut and R.~Reinbacher,
  ``The Particle spectrum of heterotic compactifications,''
  JHEP {\bf 0412} (2004) 054
  doi:10.1088/1126-6708/2004/12/054
  [hep-th/0405014].
  
\bibitem{Donagi:1999jp} 
  R.~Donagi, B.~A.~Ovrut and D.~Waldram,
  ``Moduli spaces of five-branes on elliptic Calabi-Yau threefolds,''
  JHEP {\bf 9911}, 030 (1999)
  doi:10.1088/1126-6708/1999/11/030
  [hep-th/9904054].
  
\bibitem{Braun:2004xv} 
  V.~Braun, B.~A.~Ovrut, T.~Pantev and R.~Reinbacher,
  ``Elliptic Calabi-Yau threefolds with Z(3) x Z(3) Wilson lines,''
  JHEP {\bf 0412}, 062 (2004)
  doi:10.1088/1126-6708/2004/12/062
  [hep-th/0410055].
  
\bibitem{Braun:2005ux} 
  V.~Braun, Y.~H.~He, B.~A.~Ovrut and T.~Pantev,
  ``A Heterotic standard model,''
  Phys.\ Lett.\ B {\bf 618}, 252 (2005)
  doi:10.1016/j.physletb.2005.05.007
  [hep-th/0501070].
  
\bibitem{Braun:2005bw} 
  V.~Braun, Y.~H.~He, B.~A.~Ovrut and T.~Pantev,
  ``A Standard model from the E(8) x E(8) heterotic superstring,''
  JHEP {\bf 0506}, 039 (2005)
  doi:10.1088/1126-6708/2005/06/039
  [hep-th/0502155].
  
\bibitem{Braun:2005zv} 
  V.~Braun, Y.~H.~He, B.~A.~Ovrut and T.~Pantev,
  ``Vector bundle extensions, sheaf cohomology, and the heterotic standard model,''
  Adv.\ Theor.\ Math.\ Phys.\  {\bf 10}, no. 4, 525 (2006)
  doi:10.4310/ATMP.2006.v10.n4.a3
  [hep-th/0505041].
  
\bibitem{Braun:2005nv} 
  V.~Braun, Y.~H.~He, B.~A.~Ovrut and T.~Pantev,
  ``The Exact MSSM spectrum from string theory,''
  JHEP {\bf 0605}, 043 (2006)
  doi:10.1088/1126-6708/2006/05/043
  [hep-th/0512177].
  
\bibitem{Braun:2013wr} 
  V.~Braun, Y.~H.~He and B.~A.~Ovrut,
  ``Supersymmetric Hidden Sectors for Heterotic Standard Models,''
  JHEP {\bf 1309}, 008 (2013)
  doi:10.1007/JHEP09(2013)008
  [arXiv:1301.6767 [hep-th]].
  
\bibitem{Brandle:2001ts} 
  M.~Brandle and A.~Lukas,
  ``Five-branes in heterotic brane world theories,''
  Phys.\ Rev.\ D {\bf 65}, 064024 (2002)
  doi:10.1103/PhysRevD.65.064024
  [hep-th/0109173].
  
\bibitem{Antunes:2002hn} 
  N.~D.~Antunes, E.~J.~Copeland, M.~Hindmarsh and A.~Lukas,
  ``Kinky brane worlds,''
  Phys.\ Rev.\ D {\bf 68}, 066005 (2003)
  doi:10.1103/PhysRevD.68.066005
  [hep-th/0208219].
  
 
\bibitem{Deffayet:2009wt} 
  C.~Deffayet, G.~Esposito-Farese and A.~Vikman,
  ``Covariant Galileon,''
  Phys.\ Rev.\ D {\bf 79}, 084003 (2009)
  doi:10.1103/PhysRevD.79.084003
  [arXiv:0901.1314 [hep-th]].
   
 \bibitem{Wess}
   J. ~Bagger and J.~Wess, 
   ``Supersymmetry and Supergravity'',
  Second Edition (1992),
  Princeton University Press, 
  ISBN 0-691-08556-0.
  
\bibitem{Khoury:2011da} 
  J.~Khoury, J.~L.~Lehners and B.~A.~Ovrut,
  ``Supersymmetric Galileons,''
  Phys.\ Rev.\ D {\bf 84}, 043521 (2011)
  doi:10.1103/PhysRevD.84.043521
  [arXiv:1103.0003 [hep-th]].
 
\bibitem{Koehn:2012ar} 
  M.~Koehn, J.~L.~Lehners and B.~A.~Ovrut,
  ``Higher-Derivative Chiral Superfield Actions Coupled to N=1 Supergravity,''
  Phys.\ Rev.\ D {\bf 86}, 085019 (2012)
  doi:10.1103/PhysRevD.86.085019
  [arXiv:1207.3798 [hep-th]].
  
\bibitem{Ciupke:2015msa} 
  D.~Ciupke, J.~Louis and A.~Westphal,
  ``Higher-Derivative Supergravity and Moduli Stabilization,''
  JHEP {\bf 1510}, 094 (2015)
  doi:10.1007/JHEP10(2015)094
  [arXiv:1505.03092 [hep-th]].
 
\bibitem{Baumann:2011nm} 
  D.~Baumann and D.~Green,
  ``Supergravity for Effective Theories,''
  JHEP {\bf 1203}, 001 (2012)
  doi:10.1007/JHEP03(2012)001
  [arXiv:1109.0293 [hep-th]].
  
\bibitem{Farakos:2012je} 
  F.~Farakos, C.~Germani, A.~Kehagias and E.~N.~Saridakis,
  ``A New Class of Four-Dimensional N=1 Supergravity with Non-minimal Derivative Couplings,''
  JHEP {\bf 1205}, 050 (2012)
  doi:10.1007/JHEP05(2012)050
  [arXiv:1202.3780 [hep-th]].
 
\bibitem{Farakos:2012qu} 
  F.~Farakos and A.~Kehagias,
  ``Emerging Potentials in Higher-Derivative Gauged Chiral Models Coupled to N=1 Supergravity,''
  JHEP {\bf 1211}, 077 (2012)
  doi:10.1007/JHEP11(2012)077
  [arXiv:1207.4767 [hep-th]].
  
\bibitem{Farakos:2013zya}
  F.~Farakos, C.~Germani and A.~Kehagias,
  ``On ghost-free supersymmetric galileons,''
  JHEP {\bf 1311}, 045 (2013)
  doi:10.1007/JHEP11(2013)045
  [arXiv:1306.2961 [hep-th]].

\bibitem{Ciupke:2016agp} 
  D.~Ciupke,
  ``Scalar Potential from Higher Derivative $\mathcal{N} = 1$ Superspace,''
  [arXiv:1605.00651 [hep-th]].
 
\bibitem{Hunter:2007}
 J.~D. Hunter,
 ``Matplotlib: A 2d graphics environment,"
   { Computing In Science \& Engineering}, {\bf 9}, 3 (2007)
  doi:10.1109/MCSE.2007.55

\end{thebibliography}
\end{document}